\documentclass[11pt]{revtex4}
\usepackage{amssymb}
\usepackage{latexsym}
\usepackage{epsfig}
\usepackage{float,rotating}
\usepackage{multirow}
\usepackage{mathtools}\usepackage{bbm}
\usepackage{graphicx}
\usepackage[utf8]{inputenc,xcolor}
\usepackage{graphicx}
\usepackage{array}
\usepackage{multirow,bigdelim}
\usepackage{booktabs,cancel,bigints}
\usepackage{subfigure}
\begin{document}
\title{Asymptotically (A)dS dilaton black holes with nonlinear electrodynamics}
\author{S. Hajkhalili$^{1}$\footnote{hajkhalili@gmail.com} and A. Sheykhi$^{1,2}$\footnote{asheykhi@shirazu.ac.ir}}
\address{$^1$ Physics Department and Biruni Observatory, College of
Sciences, Shiraz University, Shiraz 71454, Iran\\
$^2$ Research Institute for Astronomy and Astrophysics of Maragha
(RIAAM), P.O. Box 55134-441, Maragha, Iran}
\begin{abstract}
It is well-known that with an appropriate combination of three
Liouville-type dilaton potentials, one can construct charged
dilaton black holes in an (anti)-de Sitter [(A)dS] spaces in the
presence of linear Maxwell field. However, asymptotically (A)dS
dilaton black holes coupled to nonlinear gauge field have not been
found. In this paper, we construct, for the first time, three new
classes of dilaton black hole solutions in the presence of three
types of nonlinear electrodynamics, namely Born-Infeld,
Logarithmic and Exponential nonlinear electrodynamics. All these
solutions are asymptotically (A)dS and in the linear regime reduce
to the Einstein-Maxwell-dilaton black holes in AdS spaces. We
investigate physical properties and the causal structure, as well
as asymptotic behavior of the obtained solutions, and show that
depending on the values of the metric parameters, the singularity
can be covered by various horizons. Interestingly enough, we find
that the coupling of dilaton field and nonlinear gauge field in
the background of (A)dS spaces leads to a strange behaviour for
the electric field. We observe that the electric field is zero at
singularity and increases smoothly until reaches a maximum value,
then it decreases smoothly until goes to zero as
$r\rightarrow\infty$. The maximum value of the electric field
increases with increasing the nonlinear parameter $\beta$ or
decreasing the dilaton coupling $\alpha$ and is shifted to the
singularity in the absence of either dilaton field ($\alpha=0$) or
nonlinear gauge field ($\beta\rightarrow\infty$).\\


\end{abstract}

 \maketitle

\section{Introduction}
The investigations on the black holes solutions in the background
of AdS spacetimes have got renewed attentions in the two past
decades. This is mostly due to the Maldacena's conjecture which
corresponds a gravity theory in an AdS  space with a conformal
field theory (CFT) living on the boundary of this space, known as
AdS/CFT correspondence \cite{maldacena}. Based on this duality,
thermodynamics of black holes in $n$-dimensional AdS space can be
identified with that of a certain dual CFT in the high temperature
limit on the $(n-1)$-dimensional boundary of this spacetime. In
fact, the AdS/CFT correspondence provides an ideal tool to study
strongly coupled field theories. For example, many authors have
investigated condensed matter systems such as superconductors via
this correspondence \cite{SC}. Moreover, it was shown that a
relevance exists between thermodynamic variables and the
stress-energy tensor of large, rotating black holes and fluids on
its conformal boundary \cite{Bhattacharyya}. This fact further
motivates theoretical physicists to study black holes in the
background of (A)dS spaces.

The motivation of perusing dilaton field comes from the fact that
this field appears in the low energy limit of string theory where
the Einstein action is supplemented with other fields like axion,
gauge fields and scalar dilaton field. The dilaton field changes
the causal structure of the black hole and leads to the curvature
singularities at finite radii. In the absence of dilaton
potential, exact solutions of charged dilaton black holes have
been constructed by many authors \cite{CDB1,CDB2}. These black
holes are all asymptotically flat. It was shown that the presence
of the dilaton potential, which can be regarded as the
generalization of the cosmological constant, can change the
asymptotic behavior of the solutions to be neither flat nor (A)dS.
Indeed, it was shown that, no dilaton (A)dS black hole solution
exists with the presence of only one Liouville-type dilaton
potential \cite{MW}. In the presence of one or two Liouville-type
potential, black hole spacetimes which are neither asymptotically
flat nor (A)dS have been investigated (see e.g.
\cite{CHM,Cai1,Clem,Shey0,DHSR}). The studies where also
generalized to dilaton black holes with nonlinear electrodynamics.
Physical properties, thermodynamics and thermal stability of the
dilaton black objects in the presence of Born-Infeld (BI)
nonlinear electrodynamics have been investigated
\cite{Tam1,Tam2,YI,yaz, Clement,yazad,SRM,Shey1,Shey2}. When the
gauge field is in the forms of Exponential, Logarithmic and
Power-Maxwell nonlinear electrodynamics, dilaton black holes which
are neither asymptotically flat nor (A)dS have been constructed in
\cite{ShKa,ShN,Kord}.

It is important to note that all these solutions
(\cite{MW,CHM,Cai1,Clem,Shey0,Tam1,Tam2,YI,yaz,
Clement,yazad,SRM,Shey1,Shey2,DHSR,ShKa,Kord,ShN}), however, are
neither asymptotically flat nor (A)dS. A question then arises: Is
it possible to construct exact analytical black hole solutions of
dilaton gravity in the background of (A)dS  spaces? Gao and Zhang
were the first who answered this question and derived exact
analytical black hole solutions of Einstein-Maxwell-dilaton
gravity in the background of (A)dS spacetime \cite{gao1,gao2}. For
these purpose, they combined three Liouville-type dilaton
potentials, in an appropriate way. Following \cite{gao1,gao2}, a
lot of investigations have been done on asymptotically AdS dilaton
black holes/branes.  A class of charged rotating dilaton black
string solutions in the background of AdS spaces was presented in
\cite{SheAdS}. Thermodynamic instability of a class of
$(n+1)$-dimensional charged dilatonic spherically symmetric black
holes in the background of AdS universe have been explored in
\cite{SheAdS2}. Topological AdS black branes of
$(n+1)$-dimensional Einstein-Maxwell-dilaton theory and their
thermodynamic properties were also investigated in \cite{SheAdS3}.
Three-dimensional static and circularly symmetric solution of the
Einstein-Born-Infeld-dilaton system with (A)dS asymptotic was
constructed in \cite{Ida}. Other studies on the charged dilaton
black holes in the background of AdS spaces were carried out in
\cite{Other}.

In this paper, we would like to continue the investigations on the
asymptotically AdS dilaton black holes by considering the gauge
field in the form of nonlinear electrodynamics. As far as we know,
exact analytical (A)dS dilaton black holes coupled to nonlinear
electrodynamics in $n\geq4$ spacetime dimensions have not been
constructed. We shall consider three kind of nonlinear
electrodynamics as our gauge field, namely, Born-Infeld,
Exponential and Logarithmic nonlinear electrodynamics. Since the
asymptotic behavior of these three Lagrangian are the same as BI
case, they are well-known as the BI-type nonlinear electrodynamics
\cite{BI,Soleng,Hendi}. With the combination of three Liouville
type dilaton potentials, we are able to construct three new
classes of dilaton black hole solutions corresponding to three
type of BI-type nonlinear electrodynamics.

The organization of this  paper is as follows. In the next
section, we review the structure of the Einstein-dilaton gravity
coupled to nonlinear electrodynamics and introduce the Lagrangian
of the BI-type nonlinear electrodynamics coupled to the dilaton
field. Sections \ref{BIDsec}, \ref{EDsec} and \ref{LDsec} are
dedicated to constructing asymptotically (A)dS dilaton black hole
solutions in the presence of Born-Infeld (BI), Exponential
nonlinear (EN) and Logarithmic nonlinear (LN) electrodynamics,
respectively. In section \ref{discussion} we analyze  and discuss
the physical properties of the obtained solutions. We conclude the
paper with closing remarks in section \ref{sumsec}.
%%%%%%%%%%%%%%%%%%%%%%%%%%%%%%%%%%%%%%%%%%%%%%%%%%%%%%%%%%
\section{Action and Lagrangian}\label{Field}
We consider an action in which gravity is coupled to nonlinear
electrodynamic and dilaton field as
\begin{equation}\label{Act}
S=\frac{1}{16\pi}\int{d^{4}x\sqrt{-g}\left(\mathcal{R}\text{ }-2
g^{\mu\nu}
\partial_{\mu} \Phi \partial_{\nu}\Phi -V(\Phi
)+L(F,\Phi)\right)},
\end{equation}
where we display the Ricci scalar curvature and dilaton field with
$\mathcal{R}$ and $\Phi$, respectively. $V(\Phi)$ is the potential
for $\Phi$ and $L(F,\Phi)$ is the Lagrangian of nonlinear
electrodynamics coupled to the dilaton field given by
\cite{Shey2,ShKa,ShN}
\begin{equation}\label{lag}
L(F,\Phi)=\left\{
  \begin{array}{ll}
  $$4\beta^2e^{2\alpha\Phi}\left(1-\sqrt{1+\frac{e^{-4\alpha\Phi}F^2}{2\beta^2}}\right)\quad \quad\quad \quad $$\rm BID , &  \\&\\
    $$4\beta^{2} e^{2\alpha \Phi}\left[\exp\left(-\frac{e^{-
    4\alpha \Phi}F^2}{4\beta^{2}}\right)-1\right],\quad \quad\quad $$\rm END , &  \\
    &\\
$$ -8\beta^2 e^{2\alpha\Phi} \ln \left(1+\frac{ e^{-4\alpha\Phi}F^2}
{8 \beta^2}\right),\quad~ \quad\quad \quad$$ \rm LND,
  \end{array}
\right.
\end{equation}
The parameter $\beta$ represents the strength of the
electromagnetic field and has the dimension of mass. In fact,
$\beta$ determines the strength of the nonlinearity of the
electrodynamics. In the limit of large $\beta$
($\beta\rightarrow\infty$), the systems goes to the linear regime.
This implies that the nonlinearity of the theory disappears and
the nonlinear electrodynamic theory reduces to the linear Maxwell
electrodynamics. On the other hand, as $\beta$ decreases
($\beta\rightarrow0$), the system tends to the strong nonlinear
regime of the electromagnetic. Here, $F^2=F_{\mu \nu }F^{\mu\nu
}$, with $F_{\mu \nu }=\partial _{\mu }A_{\nu }-\partial _{\nu
}A_{\mu }$ is the electromagnetic field tensor. The constant
$\alpha$ illustrates the strength of coupling between the scalar
and electromagnetic fields. In the limit of large $\beta$,  all
these Lagrangian have similar expansion, namely
\begin{eqnarray} \label{Lag2}
&&L_{_{\rm
BID}}(F,\Phi)=-e^{-2\alpha\Phi}F^2+\frac{e^{-6\alpha\Phi}F^4}{8\beta^2}-\frac{e^{-10\alpha\Phi}F^6}{32\beta^4}+O\left(\frac{1}{\beta^6}\right),\nonumber\\\nonumber\\
&&L_{_{\rm END}}(F,\Phi)=-e^{-2\alpha\Phi}F^2+\frac{e^{-6\alpha\Phi}F^4}{8\beta^2}-\frac{e^{-10\alpha\Phi}F^6}{96\beta^4}+O\left(\frac{1}{\beta^6}\right),\nonumber\\
\nonumber\\&&L_{_{\rm
LND}}(F,\Phi)=-e^{-2\alpha\Phi}F^2+\frac{e^{-6\alpha\Phi}F^4}{16\beta^2}-\frac{e^{-10\alpha\Phi}F^6}{192\beta^4}+O\left(\frac{1}{\beta^6}\right).
\end{eqnarray}
For $\beta\rightarrow \infty$, all three BI-type Lagrangian reduce
to the standard linear Maxwell Lagrangian coupled to the dilaton
field, namely, $L=-e^{-2\alpha\Phi}F^2$ \cite{CHM}. It is
convenient to use the following simplification,
 \begin{eqnarray}
&& L_{_{\rm
BID}}(F,\Phi)=4\beta^2e^{2\alpha\Phi}\mathcal{L}(Y),~~\qquad \mathcal{L}(Y)=1-\sqrt{1+Y},\qquad ~~ Y=\frac{e^{-4\alpha\Phi}F^2}{2\beta^2},\label{BIDLag}\\[10pt]
&& L_{_{\rm END}}(F,\Phi)=4\beta^2e^{2\alpha\Phi}\mathcal{L}(Y),\qquad~~ \mathcal{L}(Y)=\exp(-Y)-1,\qquad  Y=\frac{e^{-4\alpha\Phi}F^2}{4\beta^2},\label{EDLag}\\[10pt]
&& L_{_{\rm
LND}}(F,\Phi)=-8\beta^2e^{2\alpha\Phi}\mathcal{L}(Y),\qquad
\mathcal{L}(Y)=\ln\left(1+Y\right),~~\qquad
Y=\frac{e^{-4\alpha\Phi}F^2}{8\beta^2}.\label{LDLog}
 \end{eqnarray}
In order to construct (A)dS dilaton black hole solutions, we
examine the potential  in the form \cite{gao1}
 \begin{equation}\label{pot}
 V(\Phi)={\frac {2\Lambda}{ 3\left( {\alpha}^{2}+1 \right) ^{2}} \left[8\,{\alpha}^{2}{{\rm e}^{\Phi\left(\alpha\,-{{1}/{\alpha}}\right)}}
 + {
 \alpha}^{2} \left( 3\,{\alpha}^{2}-1 \right) {{\rm e}^{-\,{ {2
 \Phi}/{\alpha}}}} - \left( {\alpha}^{2}-3 \right) {{\rm e}^{2\,\alpha\,
 \Phi}}\right] },
 \end{equation}
where $\Lambda $ is the cosmological constant. It is clear the
cosmological constant is coupled to the dilaton in a very
nontrivial way. The above potential reduces to
$V(\Phi=0)=2\Lambda$ in the absence of dilaton field
($\Phi=\alpha=0$). This type of potential can be obtained in the
cementification of the higher dimensional theory to the four
dimensions, including various super-gravity models \cite{Radu}. In
particular, by choosing $\alpha=\pm1, \pm\sqrt{1/3}, \pm\sqrt{3}$,
this potential becomes the SUSY potential \cite{gao1}.

In order to construct static and spherically symmetric black
holes, we assume the following form for the metric
\begin{equation}\label{metric}
ds^2=-f(r)dt^2+\frac{dr^2}{f(r)}+r^2
R^2(r)\Big(d\theta^2+\sin^2{\theta} d\phi^2\Big),
\end{equation}
where $f(r)$ and $R(r)$ are functions of $r$ which should be
determined. In the next three sections, we try to construct
asymptotically (A)dS dilaton black holes in the presence of three
nonlinear electrodynamics (\ref{lag}).
%%%%%%%%%%%%%%%%%%%%%%%%%%%%%%%%%%%%%%%%%%%%%%%%%%%%%%%%%%%%%%%%%%%
\section{BID black holes in (A)dS spacetime}\label{BIDsec}
In order to build asymptotically (A)dS-BID black hole, one should
obtain equations of motion by varying action (\ref{Act}) with
respect to all fields by considering the BID Lagrangian
$L(F,\Phi)$ given in (\ref{BIDLag}). The variation with respect to
the gravitational field $g_{\mu\nu}$, dilaton filed $\Phi$ and
gauge field $A_{\mu}$ yields the following field equations
 \begin{equation}\label{FEBI1}
  {\cal R}_{\mu\nu}^{^{BID}}=2 \partial _{\mu }\Phi
  \partial _{\nu }\Phi +\frac{1}{2}g_{\mu \nu }V(\Phi)-
  4e^{-2\alpha \Phi}\partial_{Y}{{\cal L}}(Y) F_{\mu\eta}
  F_{\nu}^{\text{ }\eta }+2\beta^2 e^{2\alpha \Phi}
  \left[2Y\partial_{Y}{{\cal L}}(Y)-{{\cal
  L}}(Y)\right]g_{\mu\nu},
 \end{equation}
 \begin{eqnarray}\label{FEBI2}
  \nabla ^{2}\Phi&=&\frac{1}{4}\frac{\partial V}{\partial \Phi}+
  2\alpha \beta^2 e^{2\alpha \Phi }\left[2{ Y}\partial_{Y}{{\cal
  L}}(Y)-{\cal L}(Y)\right],
 \end{eqnarray}
  \begin{equation}\label{FEBI3}
  \nabla _{\mu }\left(e^{-2\alpha \Phi}
  \partial_{Y}{{\cal L}}(Y) F^{\mu\nu}\right)=0.
  \end{equation}
Note that in case of linear electrodynamics with ${\cal
L}(Y)=-Y/2$, the system of equations (\ref{FEBI1})-(\ref{FEBI3})
reduce to the well-known equations of Einstein-Maxwell-dilaton
(EMd) gravity \cite{CHM}. First of all, the electromagnetic filed
equation (\ref{FEBI3}) can be integrated immediately, where all
components of $F_{\mu\nu}$ are zero expect $F_{tr}$,
\begin{equation}\label{FtrBI}
    F_{tr}=X^ {1/2}\beta\,{{\rm e}^{2\alpha\Phi } },
\end{equation}
where $q$ is the integration constant which is related to the
electric charge of the black hole, and
\begin{equation}
X=\frac{q^2}{{\beta}^{2}\,r^4R^4(r) +{ q}^{2}}
\end{equation}
By calculating the flux of the electromagnetic field at infinity
(Gauss's theorem), the electric charge of the black hole is
obtained as
\begin{equation}
 Q =\frac{1}{4\pi}\int e^{-2\alpha\Phi}\,{}^*F\,d \Omega
 =\frac{q \omega}{4\pi}.
\end{equation}
where $\omega=4\pi$ represents the area of a unit $2$-sphere. It
is worthwhile to note that the electric field is finite at $r=0$.
This is expected in BI theories. It is interesting to consider
three limits of Eq. (\ref{FtrBI}). First, for large $\beta$ (where
the BI action reduces to Maxwell case) we have $F_{tr}=q
e^{2\alpha \Phi}/(rR)^{2}$ as presented in \cite{CHM}. On the
other hand, if $\beta\rightarrow 0$ we get $F_{tr}=0$. Finally, in
the absence of the dilaton field ($\alpha=0$), it reduces to the
electric field of BI theory \cite{Dey,Cai2}
\begin{equation}
F_{tr}=\frac{ \beta q }{\sqrt{\beta^2 r^4+q^{2}}}.
\end{equation}
It is worth mentioning that $F_{tr}$ expression given in Eq.
(\ref{FtrBI}) for BID theory in (A)dS space, is the same as the
case of non-(A)dS space \cite{Shey2}. However,  as we shall see
shortly, the explicit form of the electric field obtained from
these two theory have complectly different behaviour. This is due
to the fact that the behavior of the dilaton field $\Phi(r)$
differs in these two cases. We will back to this issue at the end
of this section. Substituting metric (\ref{metric}) and the
electromagnetic field (\ref{FtrBI}) in the system of equations
(\ref{FEBI1}) and (\ref{FEBI2}), we can write the components of
the fields equations as
\begin{eqnarray}\label{field eq1 BI}
&&Eq_{_\Phi}=\frac{rfR}{2}\Phi^{\prime\prime}+\left(rfR^{\prime}+\frac{R}{2}
\left[2f+rf^\prime\right]\right)\phi^\prime-\frac{rR}{8}\frac{dV(\Phi)}{d\Phi}+r\alpha\beta^2R(1-\frac{1}{\sqrt{X+1}})e^{2\alpha\Phi(r)}=0,\nonumber\\,
\\&&\label{field eq2
BI}Eq_{tt}=rRf^{\prime\prime}+[2rR^{\prime}+2R]f^{\prime}+r\,R\,V(\Phi)-4rR\,
\beta^2\,e^{2\alpha\Phi}\Big(1-\frac{r^4R^4\beta^2\,X}{q^2\sqrt{X+1}}\Big)=0,
\\[15pt]&&\label{field eq3
BI}Eq_{rr}=4rf\,R^{\prime\prime}+[2rf^{\prime}+8f]R^{\prime}+\left(rf^{\prime\prime}+2f^{\prime}
+r\Big[4f{\Phi^{\prime}}^2+V(\Phi)\Big]\right)\nonumber\\&&-4rR\,\beta^2\,e^{2\alpha\Phi}\Big(1-\frac{r^4R^4\beta^2\,X}{q^2\sqrt{X+1}}\Big)=0,
\\[15pt]&&\label{field eq4
BI}Eq_{\varphi\varphi}=Eq_{\theta\theta}=\Big[rf^\prime+\frac{r^2}{2}V(\Phi)+f\Big]R^2+\Big[r^2f\,R^{\prime\prime}
+\Big(rf^\prime+4f\Big)r\,R^\prime\Big]R+r^2f\,{R^\prime}^2\nonumber\\&&~~~~~~~~~~~~~~~~~~~~-2r^2R^2\beta^2e^{2\alpha\Phi}\left(1-\frac{1}{\sqrt{X+1}}\right)-1=0,
\end{eqnarray}
where the prime indicates derivative with respect to $r$.
Eqs.(\ref{field eq1 BI})-(\ref{field eq4 BI}) contain three
unknown functions $R(r)$, $\Phi(r)$ and $f(r)$. Subtracting Eq.
(\ref{field eq2 BI}) from Eq. (\ref{field eq3 BI}), we arrive at
\begin{eqnarray}\label{Eqtt-Eqrr}
2R' +r R'' + r R \Phi'^2=0.
\end{eqnarray}
Then we make the ansatz \cite{drans}
\begin{equation}\label{ansatz}
R(r)=e^{\alpha\Phi(r)}.
\end{equation}
Inserting ansatz (\ref{ansatz}) in Eq. (\ref{Eqtt-Eqrr}), we can
find the following equation for $\Phi$
\begin{equation}\label{phi}
r \alpha \Phi''+2\alpha \Phi'+r(1+{\alpha}^{2}) \Phi'^2=0,
\end{equation}
It is easy to show that Eq. (\ref{phi}) has a solution in the form
\begin{equation}\label{phiBI}
\Phi(r)={\frac {\alpha}{{\alpha}^{2}+1}\ln  \left( 1-{\frac {b}{r}}
    \right)},\\[10pt]
\end{equation}
where  $b$ is integration constant. From (\ref{phiBI}), we see
that the solution only exist for $r\geq b$. Besides, since
$\alpha>0$, we have always $\Phi<0$, and $\Phi\rightarrow -\infty$
as $r\rightarrow b$. Further, in the asymptotic region where
$r\rightarrow\infty$ or ($r\gg b$), we have $\Phi=0$, which is an
expected result, since we are looking for the asymptotically (A)dS
solutions. The fact that we have no dilaton field at the
asymptotic region is in contrast to the case of non asymptotically
(A)dS dilaton black hole solutions \cite{Shey0}, where the dilaton
field does not vanish as $r\rightarrow\infty$ \cite{Shey2}.
Indeed, as we argued already, the presence of the dilaton field in
those cases changes the asymptotic behaviour of the solution to be
neither flat nor (A)dS. Now, we use Eq. (\ref{field eq4 BI}) to
obtain the only remaining unknown function $f(r)$. We find
\begin{eqnarray}\label{frBI}
f(r)&=&-\frac{\Lambda
}{3}r^2\left(1-\frac{b}{r}\right)^\gamma+\frac{1}{r(\alpha^2+1)-b}\left(1-\frac{b}{r}\right)
^{1-\gamma}\nonumber\\[10pt]&&\times\Bigg\{C_{1}+r(\alpha^2+1)+2\beta^2(\alpha^2+1)\int
r^2\left(1-\frac{b}{r}\right)^{2\gamma}\left(1-\sqrt{1+\eta}\right)dr\Bigg\},
\end{eqnarray}
where $C_{1}$ is a integration constant,
$\gamma=2\alpha^2/(\alpha^2+1)$ and
\begin{equation}\label{eta}
\eta=\frac{q^2}{\beta^2 r^4}\left(1-\frac{b}{r}\right)^{-2\gamma}.
\end{equation}
In the absence of dilaton field $(\alpha=0)$ the metric function
(\ref{frBI}) reduces to the metric function of BI-(A)dS black
holes \cite{Dey,Cai2}
\begin{equation} \label{alpBI0}
 f(r)=1+\frac{C_1}{r}-\frac{r^2\Lambda}{3}+\frac{2\beta^2r^2}{3}-\frac{2\beta}{r}\int\sqrt{\beta^2r^4+q^2}\,dr.
\end{equation}
For large $\beta$, one can expand (\ref{alpBI0}), to arrive at
\begin{equation} \label{frBI2}
 f(r)=1+\frac{C_1}{r}+{\frac {{q}^{2}}{{r}^{2}}}-\frac{\Lambda r^2}{3}-{\frac {{q}^{4}}{20\,{r}^{6}{
 \beta}^{2}}}+{\frac {{q}^{6}}{72\,{r}^{10}{
 \beta}^{4}}}+O \left( \frac{1}{{\beta}^{6}} \right).
\end{equation}
For $\beta\rightarrow \infty$, it has the form of charged black
hole solutions in (A)dS spaces \cite{Bril1,Cai3}. The last term in
the right-hand side of the above expression is the leading BI
correction to the topological black hole in the large $\beta$
limit. Next, we should check whether the obtained solutions
(\ref{phiBI}) and (\ref{frBI}) are satisfied all components of the
field equations or not. Our calculations show that a condition to
fully satisfy all equations of motion is required. After a number
of manipulations, we find that this condition induces a relation
between $q$, $\beta$, $C_{1}$ and the integral part of Eq.
(\ref{frBI})in the following form
\begin{eqnarray}\label{condition BI}
\frac{b+C_{1}}{2(\alpha^2+1)}-\frac{q^2}{r\,b\,\eta}\Big(r(\alpha^2+1)-b\Big)
\left[1-\sqrt{1+\eta}\right]+\beta^2\int
r^2\left(1-\frac{b}{r}\right)^{2\gamma}\left[1-\sqrt{1+\eta}\right]dr=0.
\end{eqnarray}
In the limiting case where $\beta\rightarrow \infty$,  condition
(\ref{condition BI}) reduce to the corresponding case of AdS
dilaton balck holes given in \cite{gao1}, namely
\begin{eqnarray}\label{CondBI2}
C_{1}=-\frac{b^2+(1+\alpha^2)q^2}{b}+O\left(\frac{1}{\beta^2}\right),
\end{eqnarray}
which is independent of $r$. Combining condition (\ref{condition
BI}) with solution (\ref{frBI}), we find the final form of the
solution which is fully satisfied all components of the field
equations as
\begin{eqnarray}\label{f2BI}
f(r)&=&-\frac{\Lambda r^2}{3}\left(1-\frac{b}{r}\right)^\gamma+
\Bigg{\{} 1-\frac {2(\alpha^2+1){\beta}^{2}{r}^{3}
 }{b} \left[ \sqrt {1+\eta}-1 \right]
 \left( 1-{\frac {b}{r}} \right) ^{2\gamma} \Bigg{\}}  \left( 1-{\frac {b}{r}} \right)
 ^{1-\gamma},
\end{eqnarray}
Combining $\Phi(r)$ in (\ref{phiBI}) and ansatz (\ref{ansatz})
with  relation (\ref{FtrBI}), we can obtain the explicit form of
the electric field for asymptotically (A)dS-BID black holes as
\begin{equation}\label{electricBI}
F_{tr}(r)=E(r)=\frac{q}{r^2} \frac{1}{\sqrt{1+\eta}},
\end{equation}
where $\eta$ is given by relation (\ref{eta}).  Let us note that
the electric field given in (\ref{electricBI}) differs from
electric field of BID black holes in non-(A)dS spaces given in
\cite{Shey2}. As we mentioned this is due to the difference in the
dilaton field of these two cases. Indeed, in case of
asymptotically (A)dS solution, the dilaton field vanishes as
$r\rightarrow\infty$ as it can be seen from relation
(\ref{phiBI}). While for asymptotically non-(A)dS solution, the
dilaton field diverges in the large $r$ limit \cite{Shey2}. In the
case of $\alpha=0$, electric field will be
\begin{equation}\label{EBI0}
E(r)\Big|_{\alpha=0}=\frac{q\beta}{\sqrt{q^2+\beta^2r^4}}=\frac{q}{r^2}\frac{1}{\sqrt{1+q^2/(\beta^2
r^4)}},
\end{equation}
which is finite when other metric parameters have finite values.
Also at $r=0$ it does not diverges and has a finite value
\cite{Dey}. Expanding the electric field for large $\beta$, we
arrive at
\begin{equation}\label{expandE}
E(r)=\frac{q}{r^2}-\frac{q^3}{2r^6\beta^2}\left(1-\frac{b}{r}\right)^{-2\gamma}+O\left(\frac{1}{\beta^4}\right).
\end{equation}
In the limiting case where $\beta\rightarrow\infty$ or
($r\rightarrow \infty$), the electric field does not depend on the
dilaton field and reduces to the electric field of dilaton black
holes in (A)dS spaces \cite{SheAdS}.

It is important to note that due to dilaton field (\ref{phiBI}),
our solutions do not exist for $0<r<b$ for $\alpha\neq0$. In order
to solve this problem we exclude this region from the spacetime.
One can use the following radial coordinate
\begin{equation}\label{newcoor}
r^2=\rho^2+b^2\Longrightarrow dr^2=\frac{r^2}{r^2+b^2}\,d\rho^2.
\end{equation}
So, we can introduce our metric duo to new coordinate
\begin{equation}
ds^2=-f(\rho)dt^2+\frac{\rho^2\,d\rho^2}{(\rho^2+b^2)f(\rho)}+(\rho^2+b^2)R^2(\rho)(d\theta^2+\sin^2\theta\,
d\phi^2),
\end{equation}
where $f(\rho)$ is given by Eq. (\ref{f2BI}) with replacing
$r=\sqrt{\rho^2+b^2}$. We find that the coupling of dilaton field
and nonlinear BI field in the background of (A)dS spaces leads to
a strange behaviour for the electric field.
\begin{figure}[H]
    \centering \subfigure[$\beta=0.5$]{\includegraphics[scale=0.3]{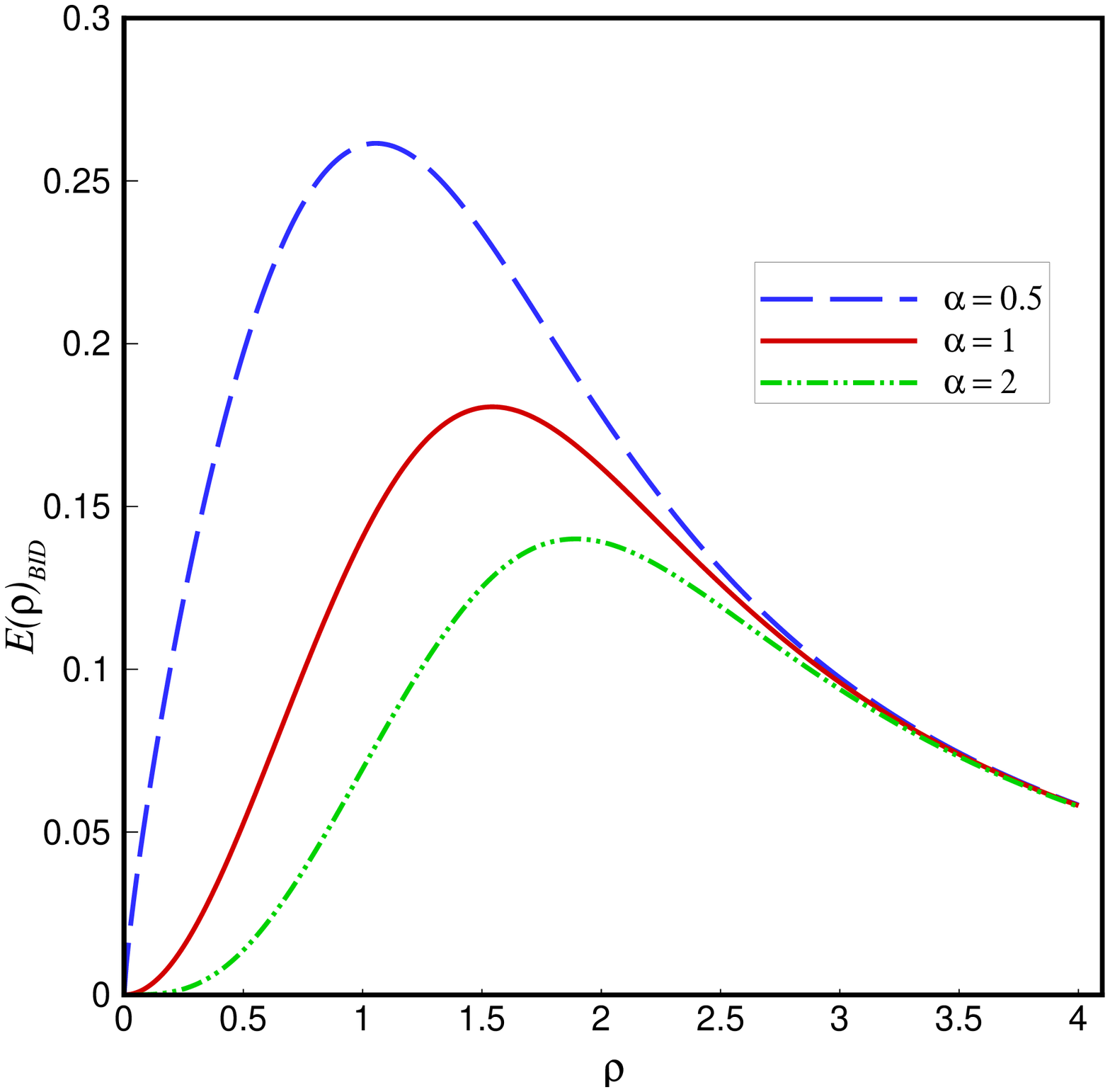}\label{fig4a}}
    \hspace*{.1cm} \subfigure[$\alpha=0.6$  ]{\includegraphics[scale=0.3]{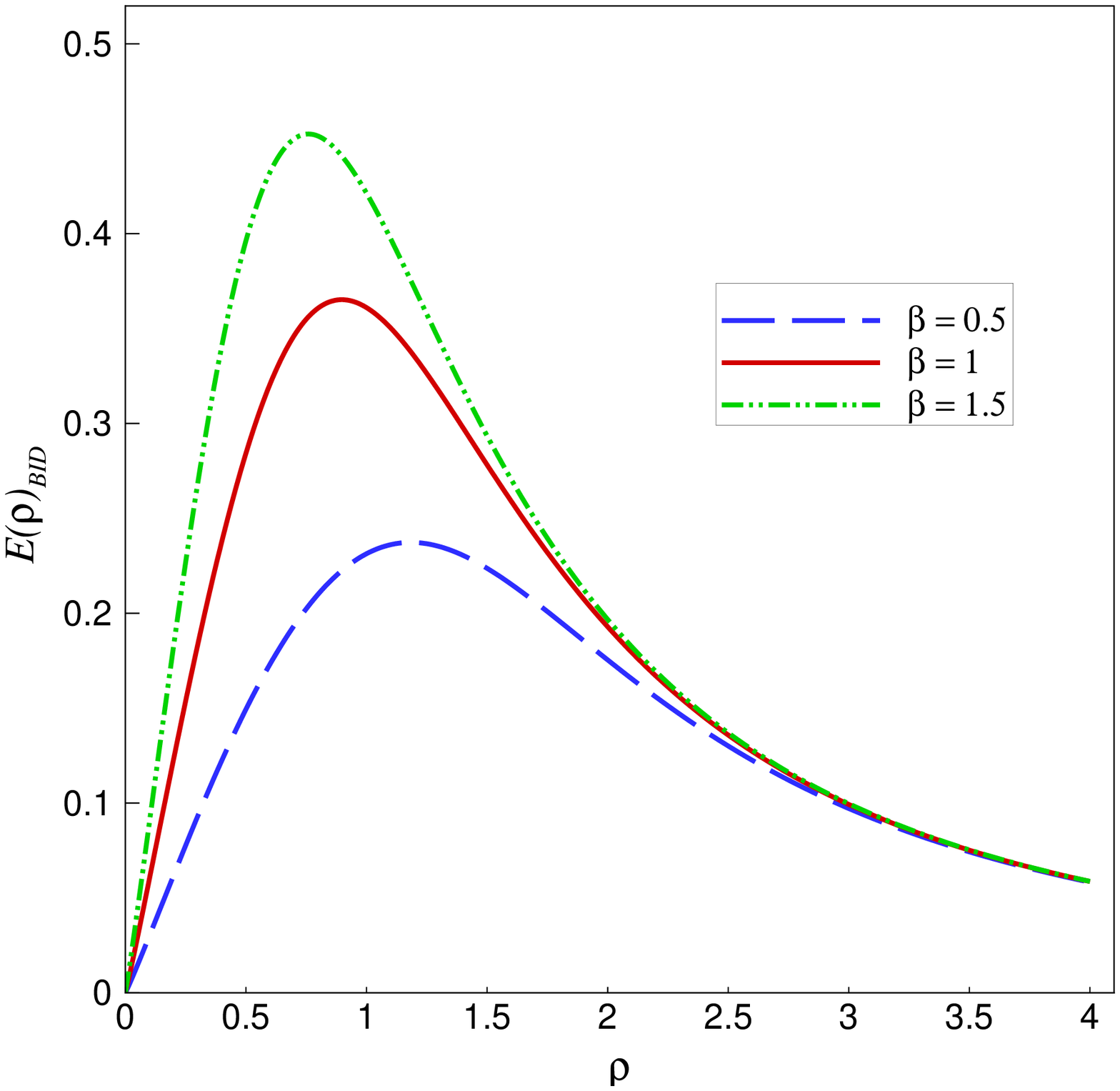}\label{fig4b}}
    \caption{The behavior electric field $E(\rho)$ for BID black holes in (A)dS spaces versus $\rho$ for $q=b=1$}\label{fig4}
\end{figure}

As one can see from Fig.\ref{fig4}, in the presence of dilaton
field ($\alpha\neq0$), the electric field has zero value at $\rho=0$.
For $\rho>0$, it increases smoothly until reaches a maximum value,
then it decreases smoothly until goes to zero as
$\rho\rightarrow\infty$. It is worth mentioning that the maximum of
the electric field increases with increasing $\beta$ or decreasing
$\alpha$. Besides, for $\alpha>0$, as  $\beta\rightarrow\infty$,
the maximum value of the electric field goes to $\rho=0$ and we have
no relative maximum for $\rho>0$. In the absence of dilaton field
($\alpha=0$), the electric field has a finite value at $\rho=0$ for a
finite value of $\beta$ as one may see from Eq. (\ref{EBI0}) that
\begin{equation}
E(\rho)\Big|_{\alpha=0,\,\rho=0}=\frac{q\beta}{\sqrt{q^2+\beta^2b^4}}
\end{equation}.
%and diverges in the Maxwell case where $\beta\rightarrow\infty$
%%%%%%%%%%%%%%%%%%%%%%%%%%%%%%%%%%%%%%%%%%%%%%%%%%%%%%%%%%%%%%%%%%%%%%%%%%%%%%%%%%%%%%%%%%%%%%%%%%%%%%%%
\section{END black holes in AdS spacetime}\label{EDsec}
In this section we examine exponential nonlinear electrodynamics
coupled to the dilaton field (END) with Lagrangian (\ref{EDLag}).
In this case the gravitational field equation is written as
\cite{ShKa}
\begin{eqnarray}\label{FEED1}
{\cal R}_{\mu\nu}&=& 2 \partial _{\mu }\Phi
 \partial _{\nu }\Phi +\frac{1}{2}g_{\mu \nu }V(\Phi)-
 2e^{-2\alpha \Phi}\partial_{Y}{{\cal L}}(Y) F_{\mu\eta}
 F_{\nu}^{\text{ }\eta }+2\beta^2 e^{2\alpha \Phi}
\left[2Y\partial_{Y}{{\cal L}}(Y)-{{\cal L}}(Y)\right]g_{\mu\nu},
  \end{eqnarray}
while the dilaton and electromagnetic field equations are still
the same as those in the BID case and given by Eqs. (\ref{FEBI2})
and (\ref{FEBI3}), respectively. Integrating the electromagnetic
field equation (\ref{FEBI3}) for END Lagrangian (\ref{EDLag}), the
only nonzero component of $F_{\mu\nu}$ is obtained as
  \begin{equation}\label{FtrED}
  F_{tr}=\frac{qe^{2\alpha\Phi(r)}}{r^2R^2(r)}\exp\left[-\frac{1}{2}L_W\left(\frac{q^2}{\beta^2r^4R^4(r)}\right)\right],
  \end{equation}
where $L_W(x)=Lambert W(x)$ is the Lambert function \cite{Lambert}
which is a special function with definition
\begin{equation*}
 L_W(x)e^{L_W(x)}=x,
\end{equation*}
where $x$ is a compelex number. It is clear that $L_W(0)=0$,
$L_W(e)=1$ and $L_W(\infty)=\infty$. The Taylor expansion of the
Lambert function is
\begin{equation}\label{lambert}
L_W(x)=\sum_{n=1}^{\infty}\frac{(-n)^{n-1}}{n!}x^n=x-x^2+\frac{3}{2}x^3-\frac{8}{3}x^4+....
\end{equation}
The series expansion (\ref{lambert}) converges for $|x|<1$. The
derivative of $L_W(x)$ can be calculated as
\begin{equation*}
L_W^\prime(x)=\frac{d
L_W(x)}{dx}=\frac{L_W(x)}{x(1+L_W(x))},\qquad\qquad\qquad \rm for
\  \   \emph{x}\neq0.
\end{equation*}
In order to get metric and dilaton functions for this type of
black holes, we use the previous ansatz (\ref{ansatz}) again. It
is a matter of calculations to show that the components of the
field equations (\ref{FEBI2}) and (\ref{FEED1})  can be written
\begin{eqnarray}\label{FEXP0}
&&Eq_{_\Phi}=\frac{rfR}{2}\Phi^{\prime\prime}+\left(rfR^{\prime}+\frac{R}{2}
\left[2f+rf^\prime\right]\right)\phi^\prime-\frac{rR}{8}\frac{dV(\Phi)}{d\Phi}+r\alpha\beta^2R\nonumber\\&&~~~~~~~~\times\left(\frac{q\big(1+2Z\big)}{r^2R^2\beta\sqrt{-2Z}}-1\right)e^{2\alpha\Phi(r)}=0,\\[15pt]&&
Eq_{tt}=rRf^{\prime\prime}+(2rR^\prime+2R)f^\prime+r\,R\,V(\Phi)
+4\beta^2rRe^{2\alpha\Phi}\left[1-\frac{q}{r^2R^2\beta}L_W\left(\frac{q^2}{\beta^2r^4R^4}\right)^{-\frac{1}{2}}\right]=0,\nonumber\\\label{FEXP1}\\[10pt]
&&Eq_{rr}=4rR^{\prime\prime}f+\Big[2rf^\prime+8f\Big]R^\prime+\Big[rf^{\prime\prime}+2f^\prime+4rf{\Phi^{\prime}}^2+r V(\Phi)\Big]R+4rR\,\beta^2e^{2\alpha\Phi}\nonumber \\
&&~~~~~~~~\times\left[1-\frac{q}{r^2R^2\beta}
L_W\left(\frac{q^2}{\beta^2r^4R^4}\right)^{-\frac{1}{2}}\right]=0,\label{FEXP2}\\[10pt]
&&Eq_{\varphi\varphi}=Eq_{\theta\theta}=\Big[rf^\prime+\frac{r^2}{2}V(\Phi)+f\Big]R^2+\left[r^2fR^{\prime\prime}
+\Big(rf^\prime+4f\Big)rR^\prime\right]R+r^2f{R^\prime}^2\nonumber \\
&&~~~~~~~~~~~~~~~~~~~~-2r^2R^2\beta^2e^{2\alpha\Phi}\Big[e^{-Z}(2Z+1)-1\Big]+1=0,
\label{FEXP3}
\end{eqnarray}
where
\begin{equation}
Z=-\frac{1}{2}L_W\left(\frac{q^2}{\beta^2r^4R^4}\right).
\end{equation}
Subtracting  Eq. (\ref{FEXP1}) from Eq. (\ref{FEXP2}), and using
the ansatz (\ref{ansatz}) we arrive at Eq. (\ref{phi}). Thus the
dilaton field is again given by Eq. (\ref{phiBI}). This implies
that the nonlinearity does not affect the dilaton field. Having
ansatz (\ref{ansatz}) and the dilaton field (\ref{phiBI}) in hand,
we can obtain the metric function $f(r)$ by solving Eq.
(\ref{FEXP3}). We find
\begin{eqnarray}\label{f1ED}
f(r)&=&-\frac{\Lambda
r^2}{3}\left(1-\frac{b}{r}\right)^\gamma+\frac{1}{r(\alpha^2+1)-b}\left(1-\frac{b}{r}\right)^{1-\gamma}
\\[10pt]&&\times\left[C_{2}+r(\alpha^2+1)-2\beta^2(\alpha^2+1)\int r^2\left(1-\frac{b}{r}\right)^{2\gamma}
\left[1+\sqrt{\eta}\left(\sqrt{L_W(\eta)}-\frac{1}{\sqrt{L_W(\eta)}}\right)\right]dr\right],\nonumber
\end{eqnarray}
where $C_{2}$ is an integration constant related to the mass of
the black hole. In the absence of dilaton field ($\alpha=0$), the
above solution reduces to
\begin{eqnarray}\label{fEXPDalp0}
f(r)&=&1+\frac{C_2}{r}-\frac{\Lambda
r^2}{3}-\frac{2\beta^2r^2}{3}-\frac{2\beta q}{r}\bigintsss
\Bigg[\sqrt{L_W\left(\frac{q^2}{r^4\beta^2}\right)}-\frac{1}{\sqrt{L_W\left(\frac{q^2}{r^4\beta^2}\right)}}\Bigg]\,dr,
\end{eqnarray}
which is the metric AdS black holes in in the presence of EN
electrodynamics \cite{AsympDrHendi}. The series expansion of
(\ref{fEXPDalp0}) for large $\beta$  becomes
\begin{equation}
f(r)=1+\frac{C_{2}}{r}+{\frac {{q}^{2}}{{r}^{2}}}-\frac{\Lambda
r^2}{3}-{\frac {{q}^{4}}{20\,{r}^{6}{
 \beta}^{2}}}+{\frac {5{q}^{6}}{216\,{r}^{10}{
\beta}^{4}}}+O \left( \frac{1}{{\beta}^{6}} \right).
\end{equation}
For $\beta\rightarrow\infty$, the above solution reduces to RN-AdS
black hole. Now, we substitute the metric function (\ref{f1ED}) in
the all components of the field equations
(\ref{FEXP0})-(\ref{FEXP3}). We find that in order to fully
satisfy all field equations, we should have the following
conditions for $q$, $\beta$ and $C_{_2}$
\begin{eqnarray}\label{condition ED}
&&\frac{b+C_{2}}{2(\alpha^2+1)}+\frac{q^2}{r\,b\,\eta}\left[1+\sqrt{\eta}\left(\sqrt{L_W(\eta)}
-\frac{1}{\sqrt{L_W(\eta)}}\right)\right]\Big(r(\alpha^2+1)-b\Big)\nonumber\\[7pt]&&-\beta^2\int r^2\left(1-\frac{b}{r}\right)^{2\gamma}
\left[1+\sqrt{\eta}\left(\sqrt{L_W(\eta)}-\frac{1}{\sqrt{L_W(\eta)}}\right)\right]dr=0.
\end{eqnarray}
It is interesting to check the limiting case of (\ref{condition
ED}) for large $\beta$,
\begin{eqnarray}\label{CondEN2}
C_{2}&=&-\frac{b^2+(1+\alpha^2)q^2}{b}+O\left(\frac{1}{\beta^2}\right),
\end{eqnarray}
which is again the condition in case of AdS dilaton balck holes
coupled to the linear Maxwell gauge field \cite{gao1}. Combining
condition (\ref{condition ED}) with solution (\ref{f1ED}), we find
the final form of the metric function, which fully satisfy the
field equations, for END black holes in the background of (A)dS
spaces as
\begin{eqnarray}\label{f2ED}
f(r)=-\frac{\Lambda r^2}{3}\left(1-\frac{b}{r}\right)^\gamma+
\Bigg{\{} 1-\frac {2q^2(\alpha^2+1)
 }{r\,b\,\eta} \left[1+\sqrt{\eta}\Bigg( \sqrt {L_W(\eta)}-\frac{1}{\sqrt{L_W(\eta)}}\Bigg) \right]
  \Bigg{\}}  \left( 1-{\frac {b}{r}} \right) ^{1-\gamma}
\end{eqnarray}
In order to calculate the explicit form of the electric field for
END-AdS black hole, we substitute the dilaton field (\ref{phiBI})
in the electromagnetics tensor (\ref{FtrED}). We find
\begin{equation}\label{EED}
E(r)=\frac{q}{r^2}\sqrt{\frac{L_W(\eta)}{\eta}}.
\end{equation}
where $\eta$ is given by (\ref{eta}). When $\alpha=0$, the
electric field of black holes in the presence of EN
electrodynamics is restored \cite{AsympDrHendi}. It is worthy to
note that the electric field (\ref{EED}) differs from the case of
non-asymptotically (A)dS dilaton black holes in the presence of EN
electrodynamics \cite{ShKa}. Also, the expansion of Eq.
(\ref{EED}) for large $\beta$, yields
\begin{equation}\label{expandED2}
E(r)=\frac{q}{r^2}-\frac{q^3}{2r^6\beta^2}\left(1-\frac{b}{r}\right)^{-2\gamma}+O\left(\frac{1}{\beta^4}\right).
\end{equation}
when $\beta\rightarrow\infty$, the electric field does not depend
on the dilaton field and reduces to the electric field of AdS
dilaton black hole \cite{SheAdS}.

\begin{figure}[H]
    \centering \subfigure[$\beta=0.5$]{\includegraphics[scale=0.3]{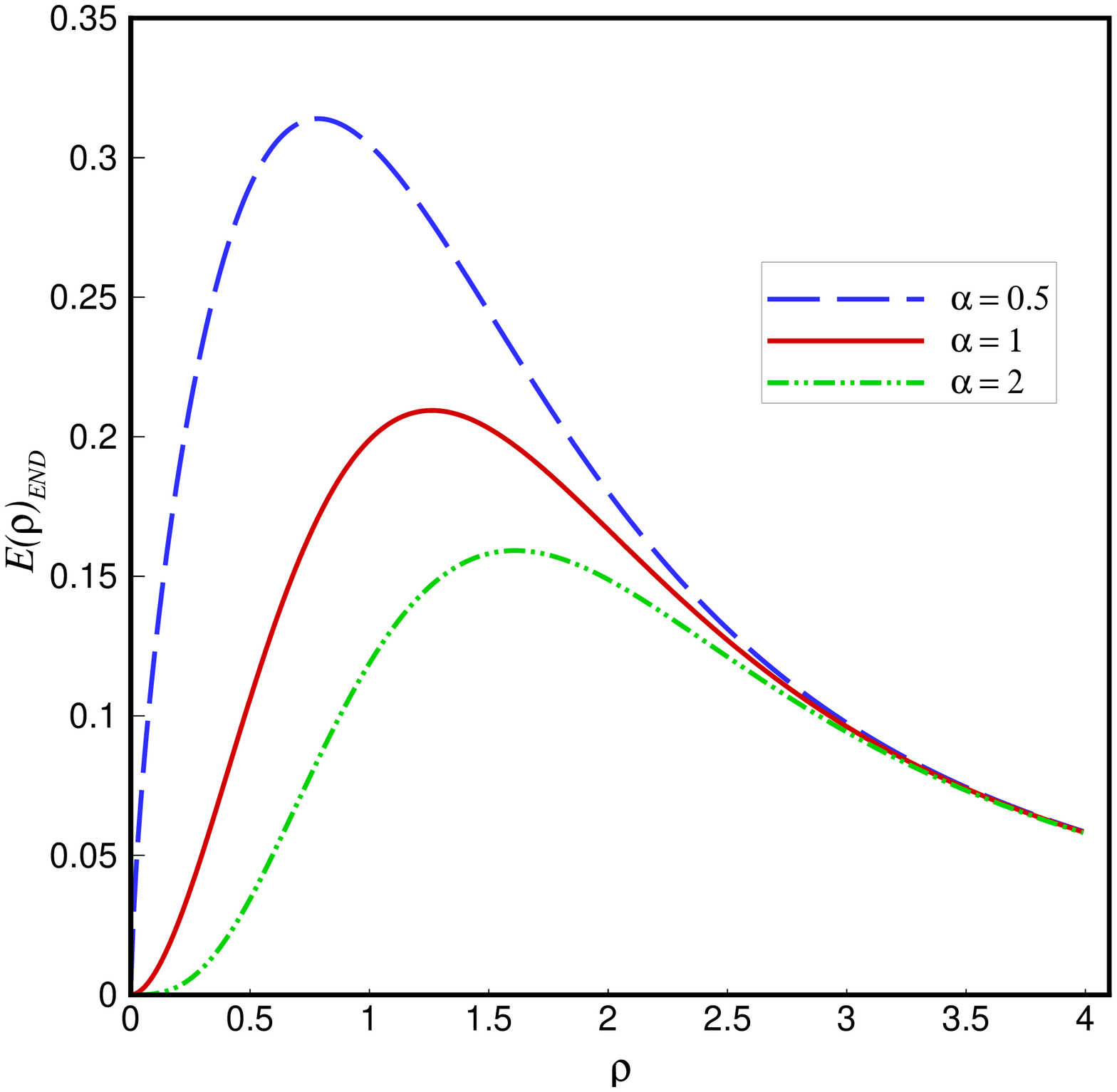}\label{fig5a}}
    \hspace*{.1cm} \subfigure[$\alpha=0.6$  ]{\includegraphics[scale=0.3]{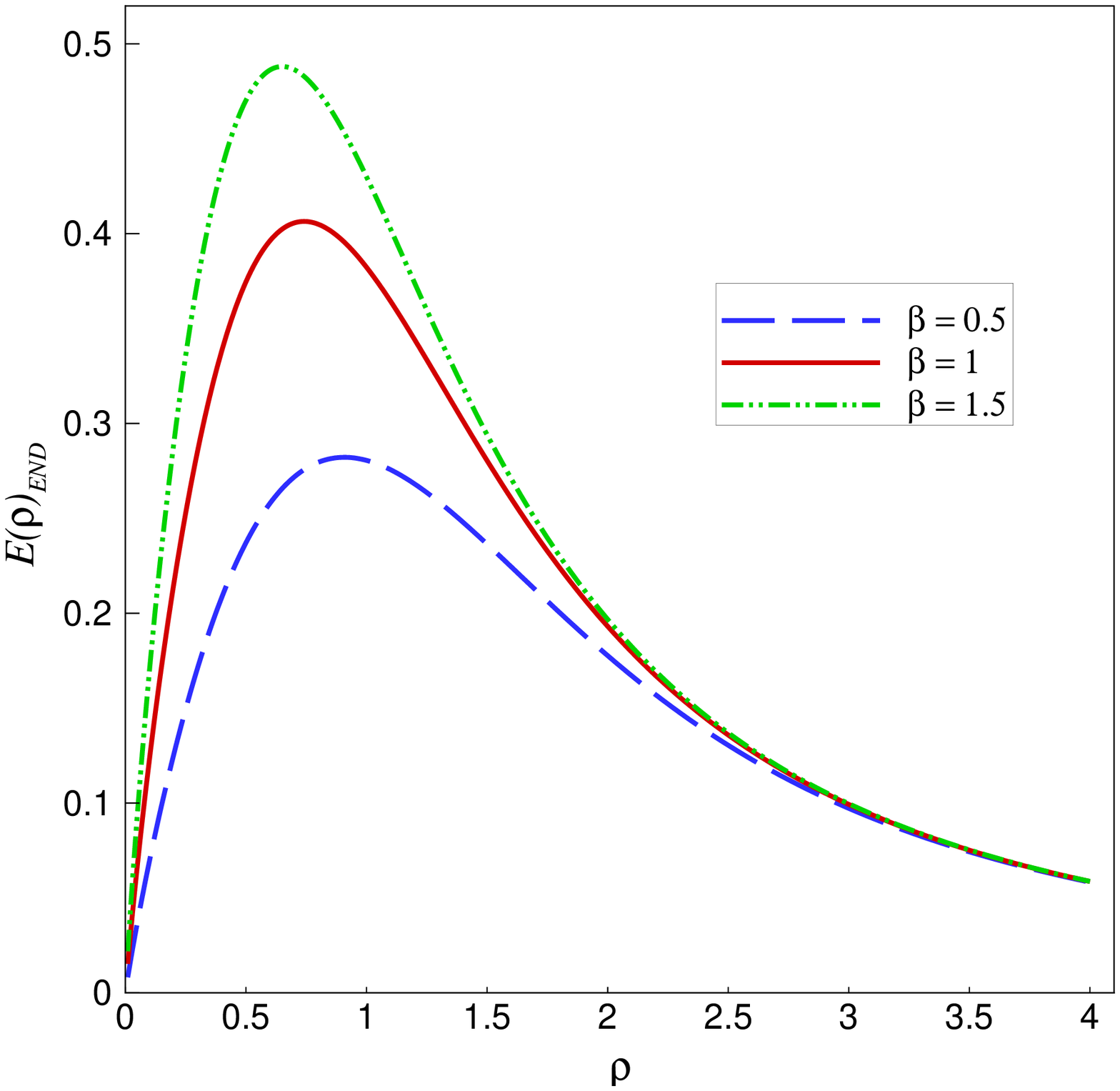}\label{fig5b}}
    \caption{The behavior
        of electric field $E(\rho)$ for END black holes in (A)dS space versus $\rho$ for $q=b=1$}\label{fig5}
\end{figure}
In order to have better understanding on the behavior of the
electric field, we plot Fig. \ref{fig5}. Again, our solution does
not exist in the range $0<r<b$, so we use the new radial
coordinate (\ref{newcoor}) to exclude this region from the
spacetime. In Fig. \ref{fig5}, $E(\rho)$ is plotted versus $\rho$
for different values of $\alpha$  and $\beta$. From this figure we
see a new behaviour for the electric field of END  black holes in
AdS spaces. As one can see the electric field vanishes at $\rho=0$
and has a maximum value at some $\rho=\rho_0>0$. The maximum point
of the electric field depends on the parameters. Indeed, for a
fixed value of $\beta$ the maximum of electric field increase as
$\alpha$ decrease (fig. \ref{fig5a}). Also, in the presence of the
dilaton field ($\alpha>0$), by increasing the nonlinear parameter
$\beta$, this maximum increases, while its position shifted to the
point $\rho_{0}\rightarrow 0$ (fig. \ref{fig5b}).
%%%%%%%%%%%%%%%%%%%%%%%%%%%%%%%%%%%%%%%%%%%%%%%%%%%%%%%%%%%%%%%%%%%%%%%%%%%%%%%%%%%%
\section{LND black holes in AdS spacetime}\label{LDsec}
In this section we would like to construct black hole solution for
Logarithmic nonlinear electrodynamics coupled to dilaton field in
the background of (A)dS spaces. Variation with respect to both
gravitational and dilaton fields yield the following field
equations \cite{ShN}
\begin{eqnarray}\label{FE1}
{\cal R}_{\mu\nu}^{LND}&=& 2 \partial _{\mu }\Phi
\partial _{\nu }\Phi +\frac{1}{2}g_{\mu \nu }V(\Phi)+2e^{-2\alpha \Phi}\partial_{Y}{{\cal L}}(Y) F_{\mu\eta}
F_{\nu}^{\text{ }\eta }-4\beta^2 e^{2\alpha \Phi}
\left[2Y\partial_{Y}{{\cal L}}(Y)-{{\cal
L}}(Y)\right]g_{\mu\nu},\qquad
\end{eqnarray}
\begin{eqnarray}\label{FE2}
 \nabla ^{2}\Phi&=&\frac{1}{4}\frac{\partial V}{\partial \Phi}-4\alpha \beta^2 e^{2\alpha \Phi }\left[2{ Y}\partial_{Y}{{\cal
 L}}(Y)-{\cal L}(Y)\right].
 \end{eqnarray}
Note that the equation of motion for the electrodynamics still
obeys Eq. (\ref{FEBI3}) with Lagrangian given in (\ref{LDLog}),
and considering  metric (\ref{metric}), one can show that it has
the following solution
\begin{equation}\label{FtrLD}
F_{tr}=\frac{2\,q\,e^{2\alpha\Phi(r)}}{r^2R^2(r)}\left(1+\sqrt{1+\frac{q^2}{\beta^2\,r^4\,R^4(r)}}\right)^{-1}.
\end{equation}
Substituting the metric (\ref{metric}) and the electric field
(\ref{FtrLD}) in the field equations (\ref{FE1}) and (\ref{FE2}), we
can write these equations as
\begin{eqnarray}\label{FELN0}
&&Eq_{_\Phi}=\frac{rfR}{2}\Phi^{\prime\prime}+\left(rfR^{\prime}+\frac{R}{2}\left[2f+rf^\prime\right]\right)\phi^\prime-\frac{rR}{8}\frac{dV(\Phi)}{d\Phi}-2r\alpha\beta^2R\nonumber\\&&~~~~~~~~\times\Bigg\{\ln\Big(1+W\Big)-\left(1+\sqrt{1+\frac{q^2}{r^4R^4\beta^2}}\right)W\Bigg\}e^{2\alpha\Phi(r)}=0\\[15pt]&&\label{FELN1}
Eq_{tt}=rRf^{\prime\prime}+\Big[2rR^\prime+2R\Big]f^\prime+r\,R\,V(\Phi)+8\beta^2rRe^{2\alpha\Phi}\nonumber \\
&&~~~~~~~~~\times
\Bigg\{\ln(W+1)-2\left(\frac{\frac{W\beta^2r^4R^4}{1+W}\sqrt{1+\frac{q^2}
{r^4R^4\beta^2}}+\frac{q^2}{2}}{\beta^2r^4R^4\sqrt{1+\frac{q^2}{r^4R^4\beta^2}}+\frac{q^2}{2}}\right)\Bigg\}=0,\\[15pt]&&\label{FELN2}
Eq_{rr}=4rR^{\prime\prime}f+\Big[2rf^\prime+8f\Big]R^\prime+\Big[rf^{\prime\prime}
+2f^\prime+4rf{\Phi^{\prime}}^2+r
V(\Phi)\Big]R-16\beta^2rRe^{2\alpha\Phi}
\nonumber\\&&~~~~~~~~~\times\Bigg\{-\frac{1}{2}\ln(W+1)
+\left(\frac{\frac{W\beta^2r^4R^4}{1+W}\sqrt{1+\frac{q^2}{r^4R^4\beta^2}}+\frac{q^2}{2}}
{\beta^2r^4R^4\sqrt{1+\frac{q^2}{r^4R^4\beta^2}}+\frac{q^2}{2}}\right)\Bigg\}=0,\\[15pt]&&\label{FELN3}
Eq_{\varphi\varphi}=Eq_{\theta\theta}=\Big[rf^\prime+\frac{r^2}{2}V(\Phi)+f\Big]R^2+\left[r^2fR^{\prime\prime}+\Big(rf^\prime+4f\Big)rR^\prime\right]R\nonumber \\
&&~~~~~~~~~~~~~~~~~~~~+r^2f{R^\prime}^2+4r^2R^2\beta^2e^{2\alpha\Phi}\left(\ln(W+1)-\frac{2W}{W+1}\right)-1=0,
\end{eqnarray}
where
\begin{equation}
W=-\frac{q^2}{r^4R^4\beta^2}\left(1+\sqrt{1+\frac{q^2}{r^4R^4\beta^2}}\right)^{-2}.
\end{equation}
Following our approach in the previous sections, we substrate Eq.
(\ref{FELN1}) from Eq. (\ref{FELN2}) and again arrive at equation
(\ref{Eqtt-Eqrr}). After using ansatz (\ref{ansatz}), we find Eq.
(\ref{phi}), which admits a solution in the form (\ref{phiBI}).
Inserting ansatz (\ref{ansatz}), the dilaton field (\ref{phiBI})
and the potential (\ref{pot}) in Eq. (\ref{FELN3}), we obtain the
following solution
\begin{eqnarray}\label{F1LD}
f(r)&=&-\frac{\Lambda
r^2}{3}\left(1-\frac{b}{r}\right)^\gamma+\frac{1}{r(\alpha^2+1)-b}
\left(1-\frac{b}{r}\right)^{1-\gamma}\nonumber\\[10pt]&&\times\Bigg\{C{_3}+r(\alpha^2+1)
-4q^2(\alpha^2+1)\int \frac{1}{r^2
\eta}\left[\frac{\eta}{1+\sqrt{1+\eta}}+\ln\left(\frac{2}{1+\sqrt{1+\eta}}\right)\right]dr\Bigg\}.
\end{eqnarray}
Here $\eta$  is again given by (\ref{eta}) and $C_{3}$ is a
constant of integration. When $\alpha=0$, the above solution
reduces to
\begin{eqnarray}\label{fEDalp0}
f(r)&=&1+\frac{C_3}{r}-\frac{\Lambda r^2}{3}-\frac{4}{r}\bigintsss
\Bigg\{\frac{q^2}{r^2}\frac{1}{1+\sqrt{1+\frac{q^2}{r^4\beta^2}}}+r^2\beta^2\ln\Bigg(\frac{2}{1+\sqrt{1+\frac{q^2}{r^4\beta^2}}}\Bigg)\Bigg\}\,dr,
\end{eqnarray}
which is the metric of  AdS black holes in in the presence of LN
electrodynamics \cite{AsympDrHendi}. The series expansion of
(\ref{fEDalp0}) for large $\beta$  becomes
\begin{equation}
f(r)=1+\frac{C_{3}}{r}+{\frac {{q}^{2}}{{r}^{2}}}-\frac{\Lambda
r^2}{3}-{\frac{{q}^{4}}{40\,{r}^{6}{\beta}^{2}}}+{\frac{{q}^{6}}{216\,{r}^{10}{\beta}^{4}}}+O \left( \frac{1}{{\beta}^{6}} \right).
\end{equation}
In the liming case where $\beta\rightarrow\infty$, the above
solution reduces to RN-AdS black hole. Inserting the metric
function (\ref{F1LD}) in the field Eqs.
(\ref{FELN1})-(\ref{FELN3}), after a time-consuming calculations,
we find a relation between constants $q$, $\beta$, $C_{3}$ and the
integration part of solution (\ref{F1LD}) as
\begin{eqnarray}\label{Cond3}
&&\frac{b+C_{3}}{4(\alpha^2+1)}+\frac{q^2}{r\,b\,\eta}\left[\frac{\eta}{1+\sqrt{1+\eta}}
+\ln\left(\frac{2}{1+\sqrt{1+\eta}}\right)
\right]\Big(r(\alpha^2+1)-b\Big)
\nonumber\\[7pt]&&-q^2 \int \frac{1}{r^2
\eta}\left[\frac{\eta}{1+\sqrt{1+\eta}}+\ln\left(\frac{2}{1+\sqrt{1+\eta}}\right)\right]dr=0,
\end{eqnarray}
Again expanding the above condition for large $\beta$, we find
\begin{eqnarray}\label{CondLN2}
C_{3}=-\frac{b^2+(1+\alpha^2)q^2}{b}+O\left(\frac{1}{\beta^2}\right),
\end{eqnarray}
Combining condition (\ref{Cond3}) with solution (\ref{F1LD}), we
arrive at the final form of the metric function for LND black
holes in (A)dS spaces as
 \begin{eqnarray}\label{f2LD}
 f(r)&=&-\frac{\Lambda r^2}{3}\left(1-\frac{b}{r}\right)^\gamma+ \Bigg{\{} 1-\frac {4q^2(\alpha^2+1)
  }{r\,b\,\eta} \left[\frac{\eta}{1+\sqrt{1+\eta}}+\ln\left(\frac{2}{1+\sqrt{1+\eta}}\right) \right]
\Bigg{\}}  \left( 1-{\frac {b}{r}} \right) ^{1-\gamma}.
 \end{eqnarray}
Now we back to the calculate the explicit form of the electric
field. One can show that the combination of dilaton field, Eq.
(\ref{ansatz}) and Eq. (\ref{FtrLD}) results the electric field
for this type of black hole as
\begin{equation}
E(r)=\frac{2\,q}{r^2}\frac{1}{1+\sqrt{1+\eta}}.
\end{equation}
In the absence of dilaton field ($\alpha=0$), this expression for
the electric field reduces to one presented in
\cite{AsympDrHendi}. Let us note that in the large limit of
nonlinear parameter $\beta$, the electric field becomes
 \begin{equation}
 E(r)=\frac{q}{r^2}+\frac{q^3}{4\,r^6\,\beta^2}\left(1-\frac{b}{r}\right)^{2\gamma}+O\left(\frac{1}{\beta^4}\right).
\end{equation}
It is clear that the first term in the above relation is the
electric field of  EMd black holes in AdS spaces \cite{SheAdS}
which is independent of dilaton coupling $\alpha$, and the second
term is the leading order correction term which appears due to the
nonlinearity of the electrodynamic field.

\begin{figure}[H]
    \centering \subfigure[$\beta=0.5$]{\includegraphics[scale=0.3]{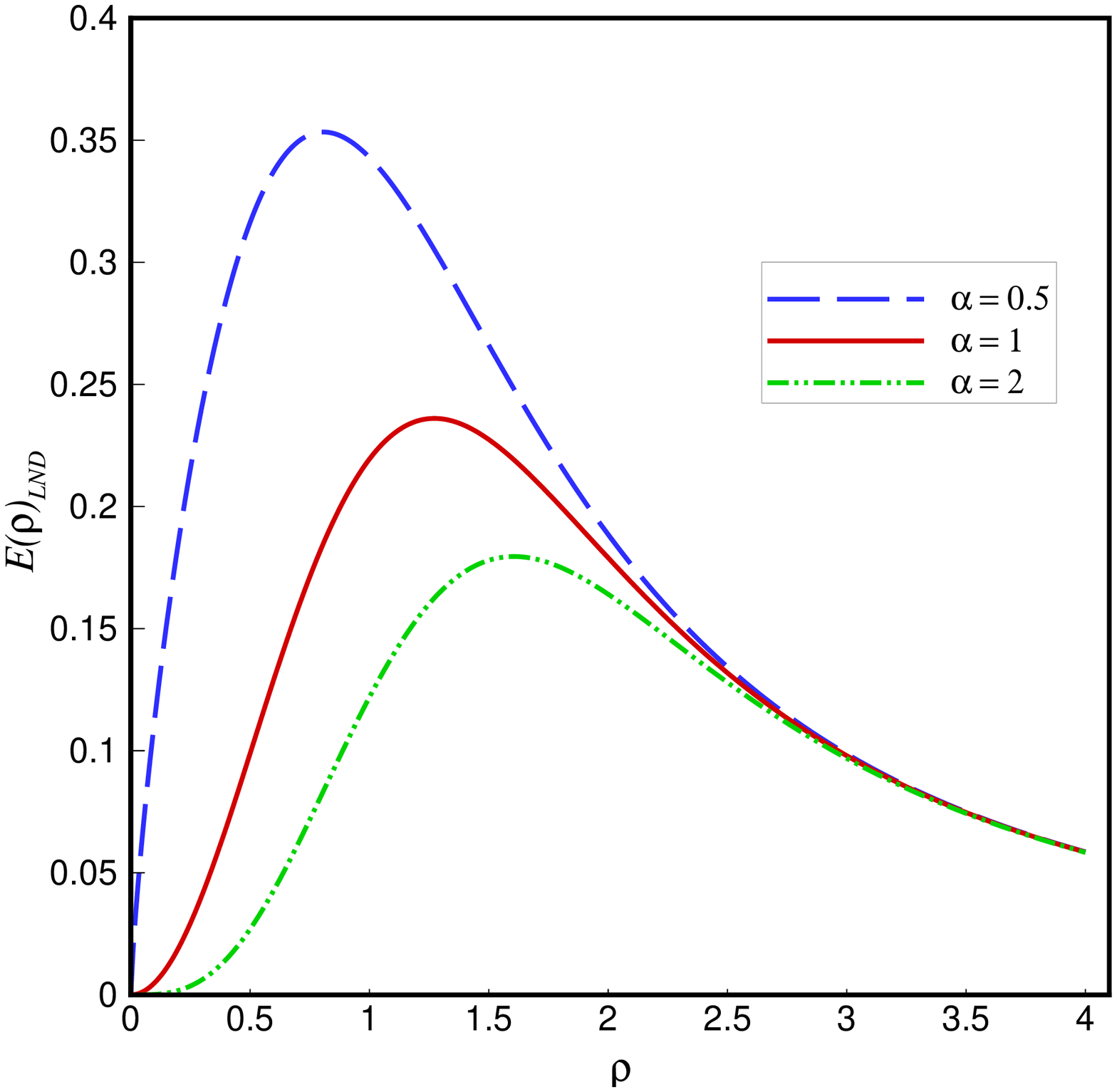}\label{fig6a}}
    \hspace*{.1cm} \subfigure[$\alpha=0.6$  ]{\includegraphics[scale=0.3]{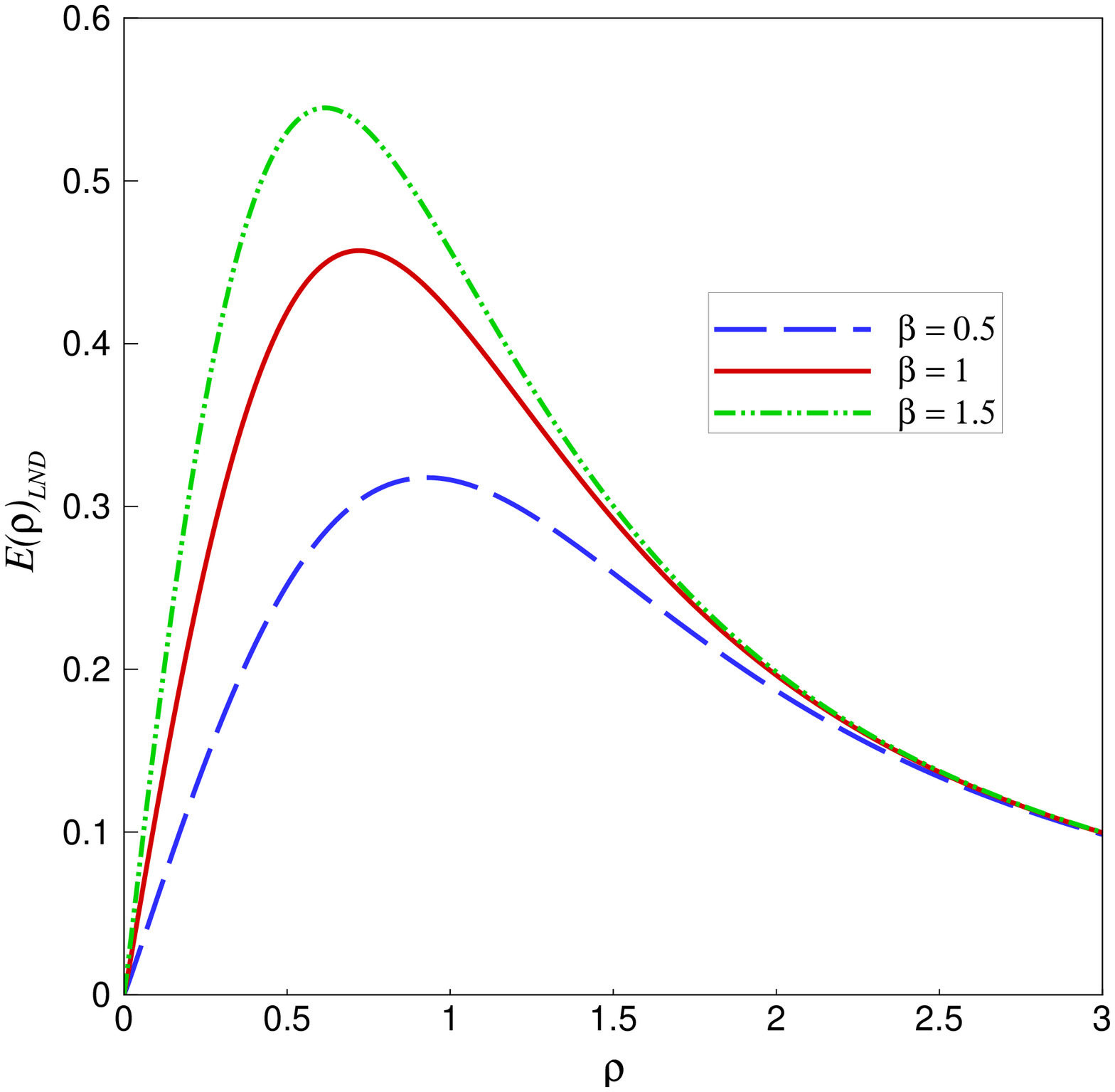}\label{fig6b}}
    \caption{The behavior
        of $E(\rho)$ for LND black holes in (A)dS space versus $\rho$ for $q=b=1$}\label{fig6}
\end{figure}

Let us note that at $\rho=0$ the electric field becomes zero like
the two previous nonlinear electrodynamics. The behavior of the
electric field respect to $\rho$ is plotted in Fig. \ref{fig6}.
From this figure, we observe that in the presence of the dilaton
field, the behaviour of the electric field is similar to the
electric field of END and BID black holes in the previous
sections.
%%%%%%%%%%%%%%%%%%%%%%%%%%%%%%%%%%%%%%%%%%%%%%%%%%%%%%%%%%%
\section{Discussion}\label{discussion}
Now, we want to study the casual structure and physical properties
of the obtained solutions. Let us remember that the metric
function of asymptotically (Ad)S black holes in EMd gravity has
the following form \cite{SheAdS}
 \begin{eqnarray}\label{fMD}
 f(r)_{\rm EMd}=\left(1-\frac {c
  }{r} \right)  \left( 1-{\frac {b}{r}} \right) ^{1-\gamma}-\frac{\Lambda
  r^2}{3}\left(1-\frac{b}{r}\right)^\gamma,
 \end{eqnarray}
where the two constants $c$ and $b$ are satisfied
$q^2(\alpha^2+1)=bc$. Expanding the obtained solutions in Eqs.
(\ref{f2BI}), (\ref{f2ED}) and (\ref{f2LD}) for large $\beta$, we
find
\begin{eqnarray}
f(r)_{\rm BID}&=&f(r)_{\rm END}=f(r)_{\rm EMd}+\frac{q^4(\alpha^2+1)}{4\,br^5\beta^2}\left( 1-{\frac {b}{r}} \right) ^{1-3\gamma}+O\left(\frac{1}{\beta^4}\right)\nonumber\\[20pt]&&
f(r)_{\rm LND}=f(r)_{\rm EMd}+\frac{q^4(\alpha^2+1)}{8\,br^5\beta^2}\left( 1-{\frac {b}{r}} \right) ^{1-3\gamma}+O\left(\frac{1}{\beta^4}\right)
\end{eqnarray}
It is obvious that for $r\rightarrow\infty$, the dominant term in
all solutions is $-{\Lambda r^2}/{3}$, which guarantees that the
asymptotic behavior of the metric function is (A)dS for
($\Lambda<0$) $\Lambda>0$. The expansion of these solutions in the
absence of dilaton field ($\alpha=0$) and for $r\rightarrow\infty$
can be written as
 \begin{eqnarray}
&& f(r)_{\rm BID}=f(r)_{\rm END}=  -\frac{\Lambda r^2}{3}+\left( 1-{\frac {b}{r}}
\right)\Bigg\{1-\frac {c
}{r}+\frac{q^4}{4br^5\beta^2}-\frac{q^6}{8br^9\beta^4}\Bigg\}
+O\left(\frac{1}{\beta^{6}}\right),\nonumber \\
&&f(r)_{\rm LND}=-\frac{\Lambda r^2}{3}+\left( 1-{\frac {b}{r}}
\right)\Bigg\{1-\frac {c
}{r}+\frac{q^4}{8br^5\beta^2}-\frac{q^6}{24br^9\beta^4}\Bigg\}+O\left(\frac{1}{\beta^{6}}\right).
\end{eqnarray}
Now, we are going to investigate the geometric nature of the
solutions. The existence of singularity and horizon ensure that
solutions are intrinsically black holes. To cover all the
spacetime, we prefer to continue with the new coordinate $\rho$.
One can compute the Ricci and Kretschmann scalars as
\begin{eqnarray}
Ricci&=&-\left(1+\frac{b^2}{\rho^2}\right)f^{\prime\prime}+\Bigg[-4\rho(b^2+\rho^2)R^{\prime}+(b^2-4\rho^2)R\Bigg]\frac{f^\prime}{R\rho^3}+\frac{2}{R^2(b^2+\rho^2)}\nonumber\\&&+\Bigg[-4\rho (b^2+\rho^2)R^{\prime\prime}R-2\rho(b^2+\rho^2){R^{\prime}}^2+4(b^2-3\rho^2)R^\prime R-\frac{2\rho^3R^2}{b^2+\rho^2}\Bigg]\frac{f}{R^2\rho^3}\nonumber\\[10pt]
R_{\mu\nu\rho\sigma}R^{\mu\nu\rho\sigma}&=&\frac{(b^2+\rho^2)^2}{\rho^4
}{f^{\prime\prime}}^2+\Bigg[4(b^2+\rho^2)^2{R^{\prime}}^2+8\rho(b^2+\rho^2)
RR^\prime+\frac{(b^4+4\rho^4)R^2}{\rho^2}\Bigg]\frac{{f^{\prime}}^2}{\rho^4 R^2}\nonumber\\&&
+\Bigg\{-\frac{2b^2(b^2+\rho^2)f^{\prime\prime}}{\rho^5}+8\Big[\rho(b^2+\rho^2)
R^{\prime\prime}-(b^2-2\rho^2)R^\prime\Big]\
\Big((b^2+\rho^2)R^\prime+\rho R\Big)\frac{f}{\rho^5
    R^2}\Bigg{\}}f^\prime\nonumber\\&&+\Bigg\{\frac{4\rho(b^2+\rho^2)^2{R^{\
            \prime}}^4}{R}+16\rho^3(b^2+\rho^2){R^{\prime}}^3+\frac{16\rho^5R^2R^\prime}{(b^2+\rho^2)}+\frac{4\rho^6R^3}{(b^2+\rho^2)^2}\nonumber\\&&+8\Bigg[\rho^2(b^2+\rho^2)^2{R^{\prime\prime}}^2-2\rho(b^2+\rho^2)(b^2-2\rho^2)R^\prime R^{\prime\prime}+(b^4-4b^2\rho^2+7\rho^4){R^{\prime}}^2\Bigg]R\Bigg{\}}\frac{f^2}{
    \rho^6R^3}\nonumber\\&&+\left(-8(b^2+\rho^2){R^\prime}^2-16\rho RR^\prime-\frac{8\rho^2R^2}{b^2+\rho^2}\right)\frac{f}{(b^2+\rho^2)\rho^2R^4}+\frac{4}{(b^2+\rho^2)^2R^4}
\end{eqnarray}
where the prime indicates the derivative with respect to $\rho$.
Also, $R=R(\rho)$ and $f=f(\rho)$ where may find them as
\begin{eqnarray}
R(\rho)&=&\left(1-\frac{b}{\sqrt{b^2+\rho^2}}\right)^{\frac{\gamma}{2}}\\
f(\rho)_{BID}&=&-\left(1-\frac{b}{\sqrt{b^2+\rho^2}}\right)^\gamma\frac{(b^2+\rho^2)\Lambda}{3}+\left(1-\frac{b}{\sqrt{b^2+\rho^2}}\right)^{1-\gamma}\nonumber\\&&
\times\Bigg{\{}1-\frac{2\beta^2(\alpha^2+1)}{b}(b^2+\rho^2)^{\frac{3}{2}}\left(1-\frac{b}{\sqrt{b^2+\rho^2}}\right)^{2\gamma}\left[\sqrt{1+\eta_\rho}-1\right]\Bigg{\}}\\
f(\rho)_{END}&=&
-\left(1-\frac{b}{\sqrt{b^2+\rho^2}}\right)^\gamma\frac{(b^2+\rho^2)\Lambda}{3}+\left(1-\frac{b}{\sqrt{b^2+\rho^2}}\right)^{1-\gamma}\nonumber\\&&
\times
\Bigg{\{} 1-\frac {2q^2(\alpha^2+1)
}{\,b\,\eta_\rho\sqrt{b^2+\rho^2}} \left[1+\sqrt{\eta_\rho}\Bigg( \sqrt {L_W(\eta_\rho)}-\frac{1}{\sqrt{L_W(\eta_\rho)}}\Bigg) \right]
\Bigg{\}} \\
f(\rho)_{LND}&=&-\left(1-\frac{b}{\sqrt{b^2+\rho^2}}\right)^\gamma\frac{(b^2+\rho^2)\Lambda}{3}+\left(1-\frac{b}{\sqrt{b^2+\rho^2}}\right)^{1-\gamma}\nonumber\\&&
\times
\Bigg{\{} 1-\frac {4q^2(\alpha^2+1)
}{b\,\eta_\rho\sqrt{b^2+\rho^2}} \left[\frac{\eta_\rho}{1+\sqrt{1+\eta_\rho}}+\ln\left(\frac{2}{1+\sqrt{1+\eta_\rho}}\right) \right]
\Bigg{\}}
\end{eqnarray}
where
\begin{equation}
\eta_\rho=\frac{q^2}{\beta^2(b^2+\rho^2)^2}\left(1-\frac{b}{\sqrt{b^2+\rho^2}}\right)^{-2\gamma}
\end{equation}
It is a matter of calculation to show that these scalars diverge
in the vicinity of the origin, namely
\begin{eqnarray}
\lim\limits_{\rho\rightarrow0}Ricci&=&\infty,\nonumber\\\lim\limits_{\rho\rightarrow0}R_{\mu\nu\rho\sigma}R^{\mu\nu\rho\sigma}&=&\infty.\nonumber
\end{eqnarray}
These confirm that the spacetime associated with the solutions has
a singularity at the origin ($\rho=0$). Also, it can be shown that
in large distances and for various values of $\alpha$, the Ricci
and Kretschmann scalars are $N\Lambda$, where $N$ is a positive
number, so it guarantees that the asymptotical behavior of these
solutions is (A)dS for $\Lambda<0$ ($\Lambda>0$). Now, we focus on
the existence of horizon.

In order to find the location of all horizons one should solve
$f(\rho=\rho_{+})$=0. Due to the complexity of the obtained metric
functions, it is not possible to solve this equation analytically.
To have better understanding on the nature of horizons, we plot
$f(\rho)$ versus $\rho$ for different metric parameters.
 \begin{figure}[H]
    \centering \subfigure[$f(\rho)_{_{\rm BID}}$ for $q=0.97$]{\includegraphics[scale=0.3]{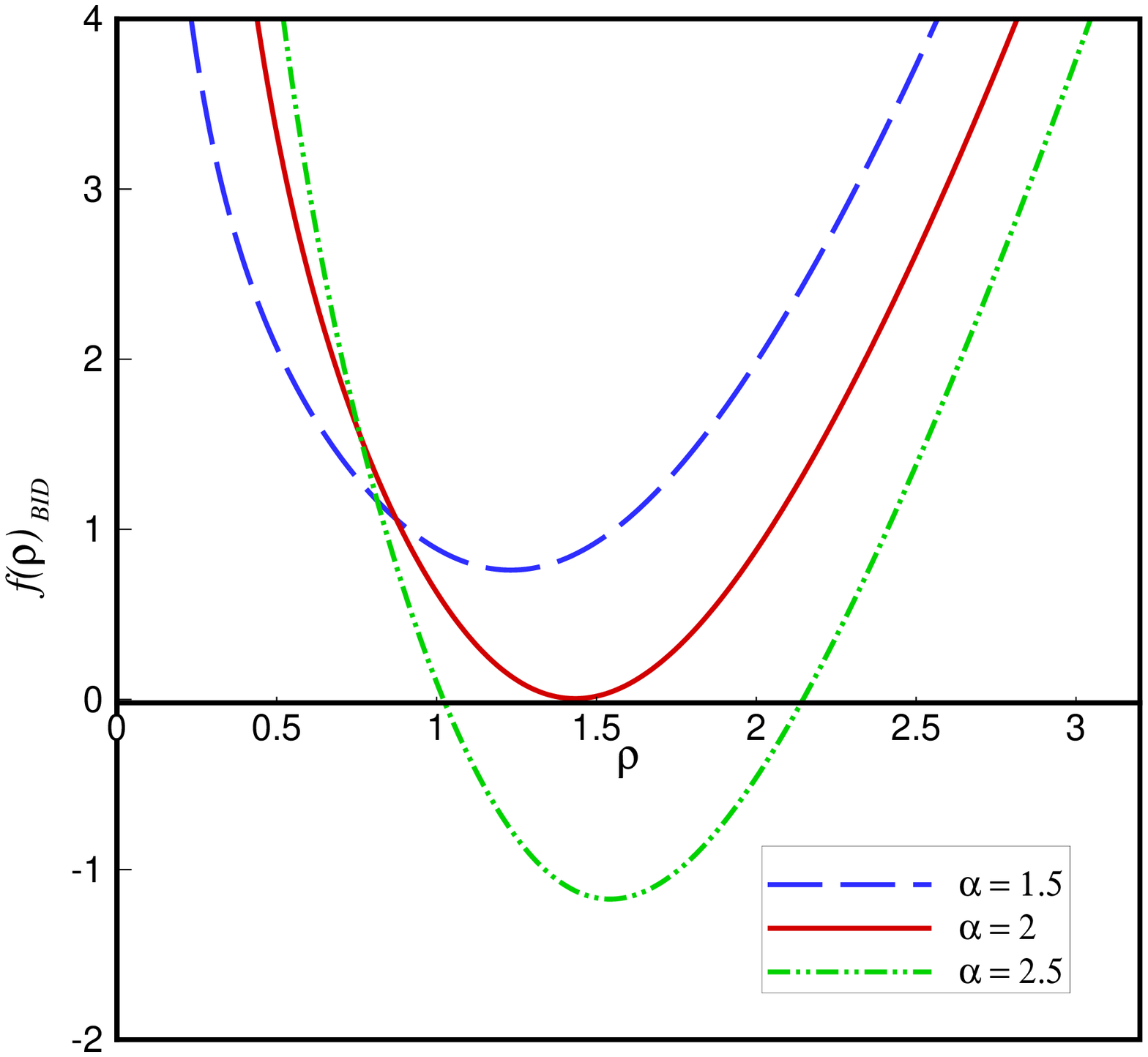}\label{fig1a}}
    \hspace*{.1cm} \subfigure[$f(\rho)_{_{\rm END}}$ for $q=0.87$  ]{\includegraphics[scale=0.3]{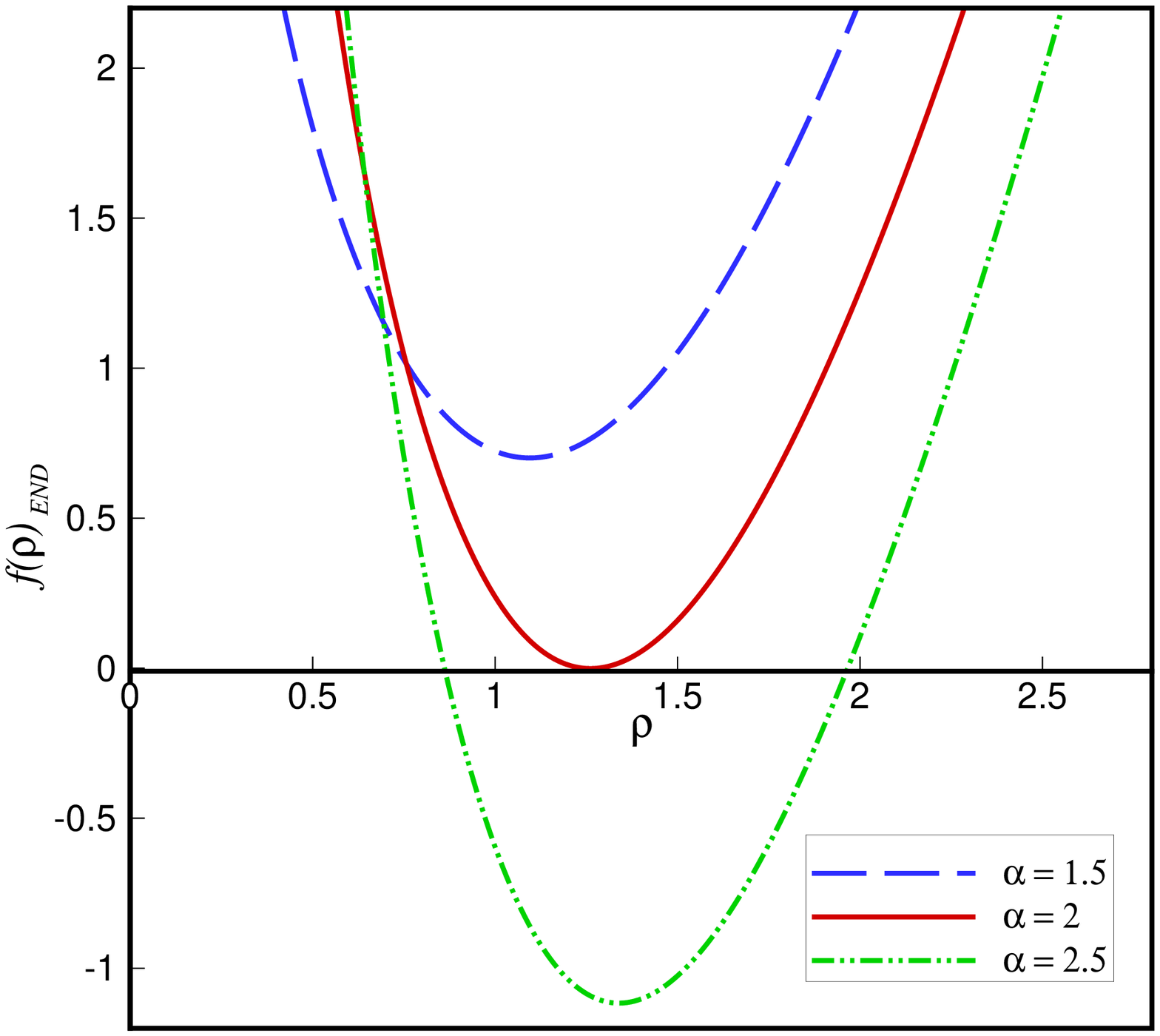}\label{fig1b}}
    \hspace*{.1cm} \subfigure[$f(\rho)_{_{\rm LND}}$ for $q=0.81$  ]{\includegraphics[scale=0.3]{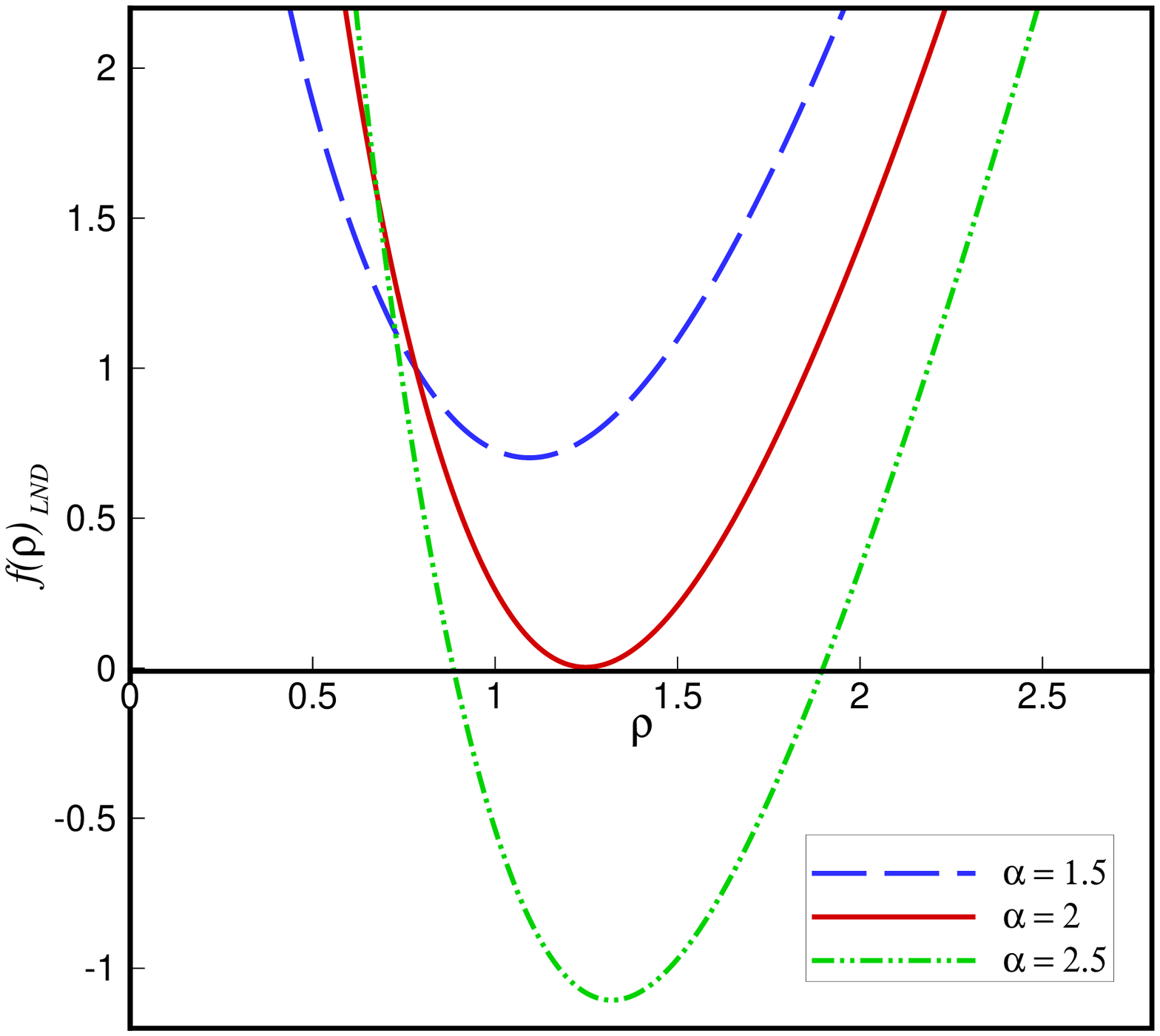}\label{fig1c}}\caption{The behavior
        of $f(\rho)$ versus $\rho$ for $\beta=0.5$, $\Lambda=-3$ and $b=1$.}\label{fig1}
\end{figure}
\begin{figure}[h]
    \centering \subfigure[$f(\rho)_{_{\rm BID}}$ for $q=1.13$]{\includegraphics[scale=0.3]{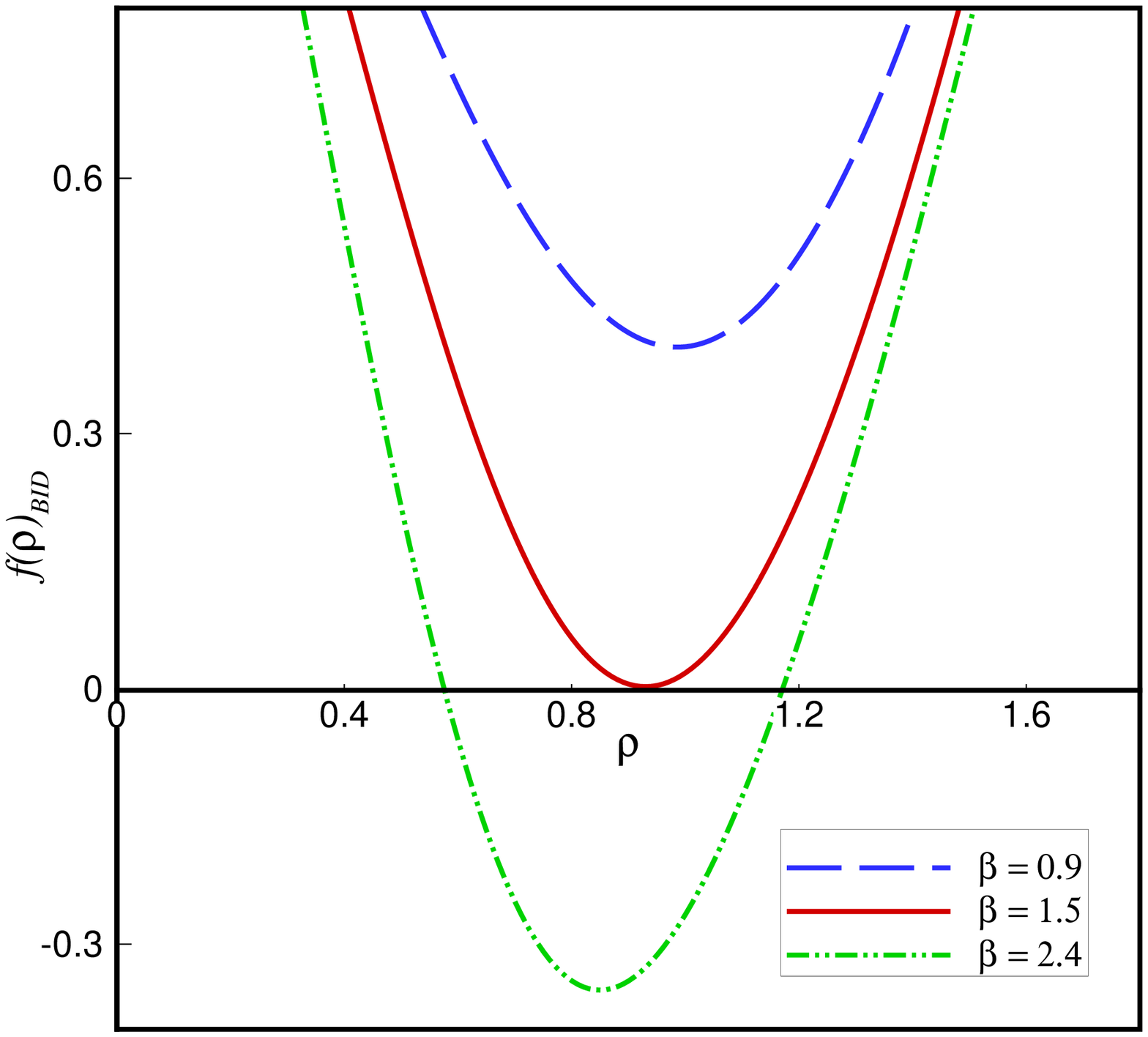}\label{fig2a}}
    \hspace*{.1cm} \subfigure[$f(\rho)_{_{\rm END}}$ for $q=1.064$  ]{\includegraphics[scale=0.3]{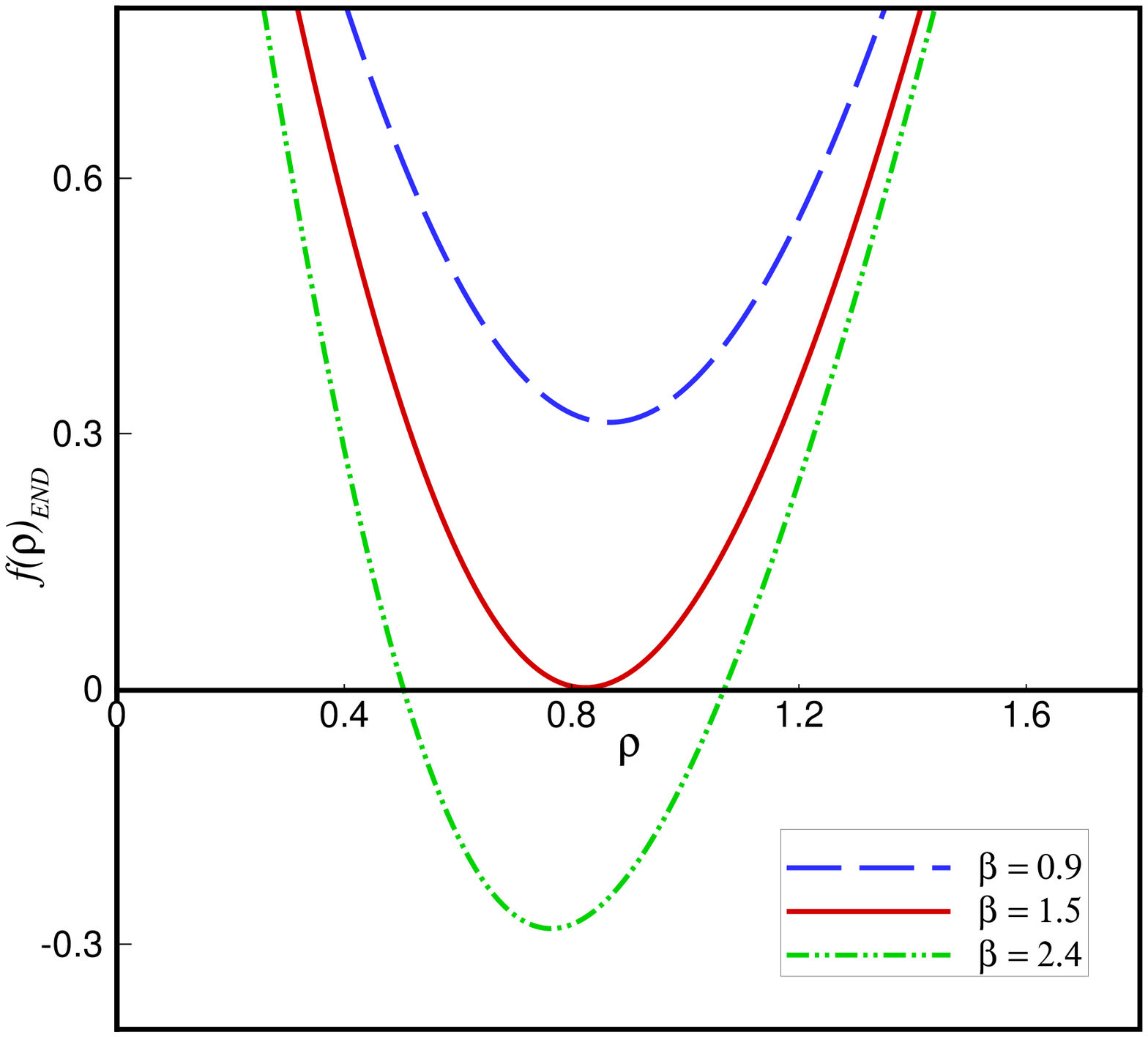}\label{fig2b}}
    \hspace*{.1cm} \subfigure[$f(\rho)_{_{\rm LND}}$ for $q=1$  ]{\includegraphics[scale=0.3]{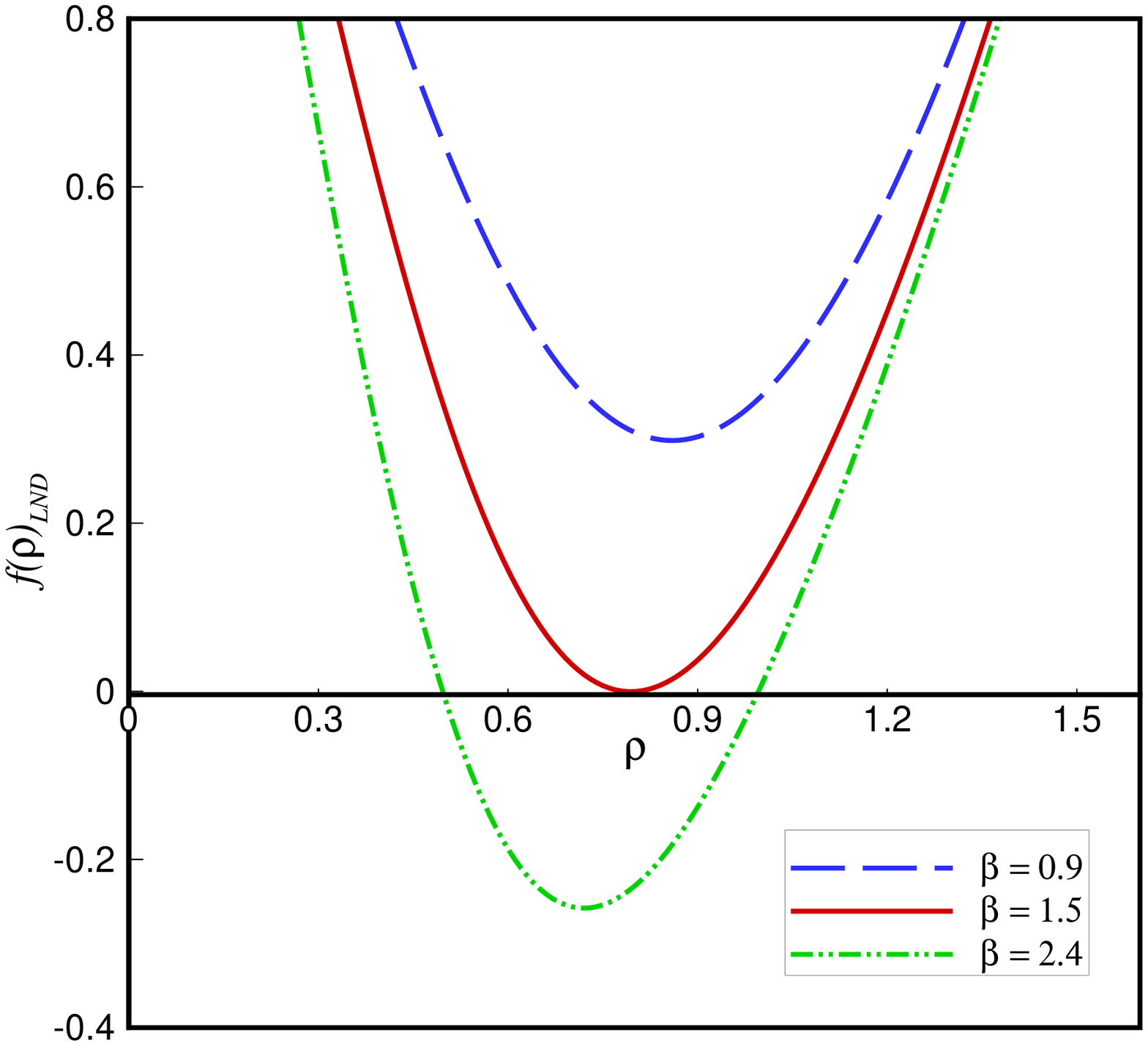}\label{fig2c}}\caption{The behavior
        of $f(\rho)$ versus $\rho$ for $\alpha=1.1$, $\Lambda=-3$ and $b=1$.}\label{fig2}
\end{figure}
\begin{figure}[h]
    \centering \subfigure[$f(\rho)_{_{\rm BID}}$ for $\beta=0.47$]{\includegraphics[scale=0.3]{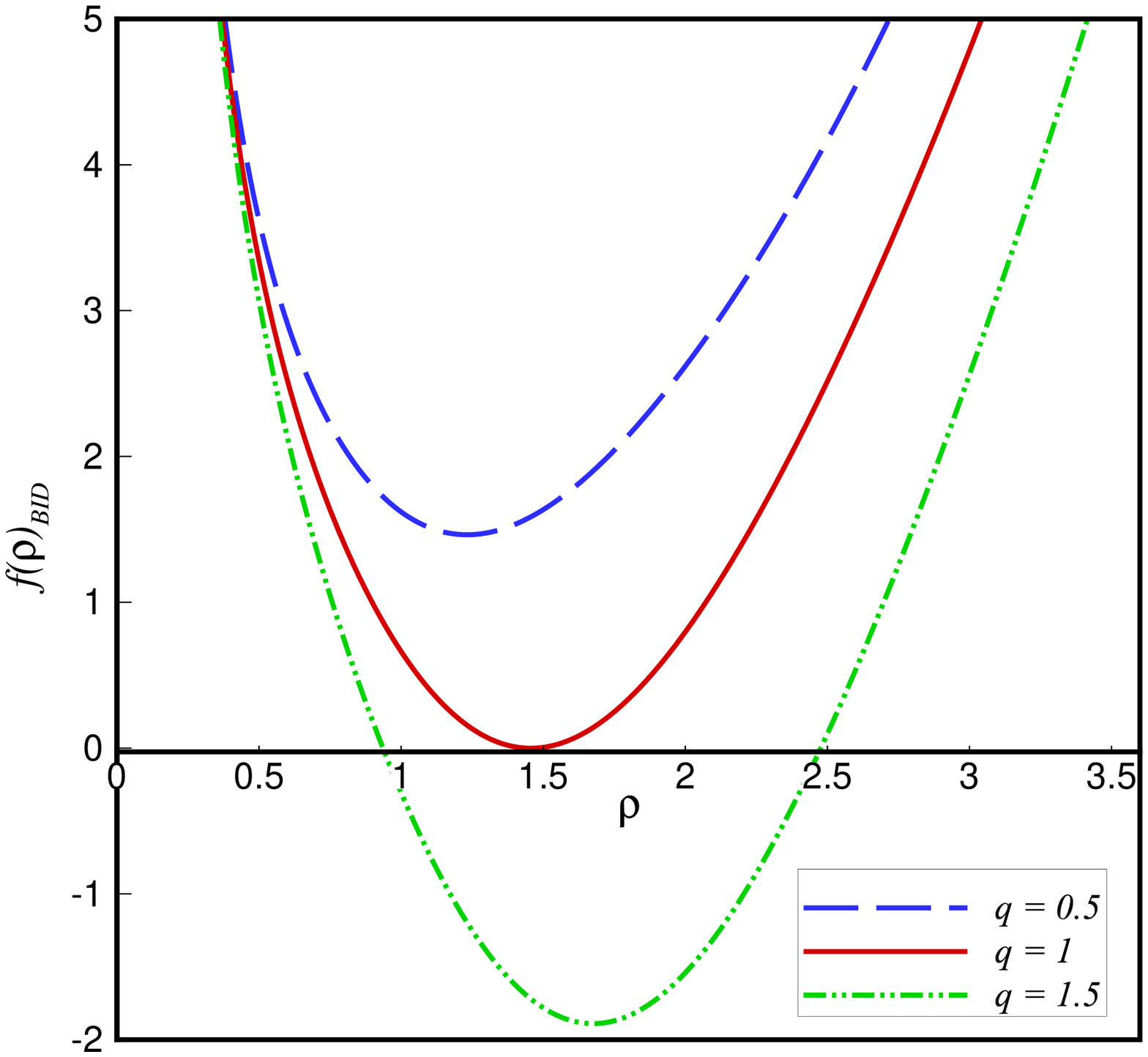}\label{fig3a}}
    \hspace*{.1cm} \subfigure[$f(\rho)_{_{\rm END}}$ for $\beta=0.33$  ]{\includegraphics[scale=0.3]{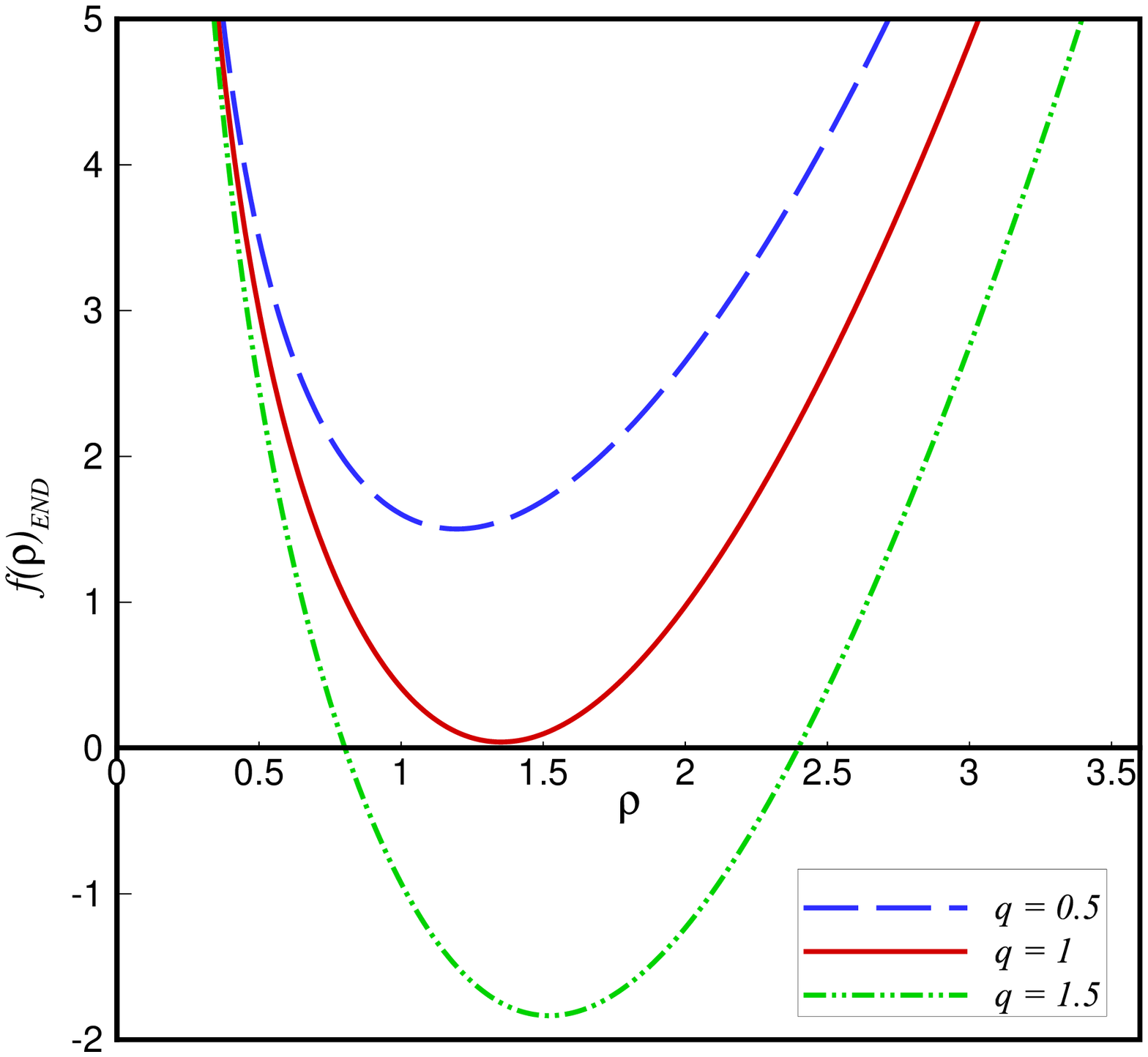}\label{fig3b}}
    \hspace*{.1cm} \subfigure[$f(\rho)_{_{\rm LND}}$ for $\beta=0.29$  ]{\includegraphics[scale=0.3]{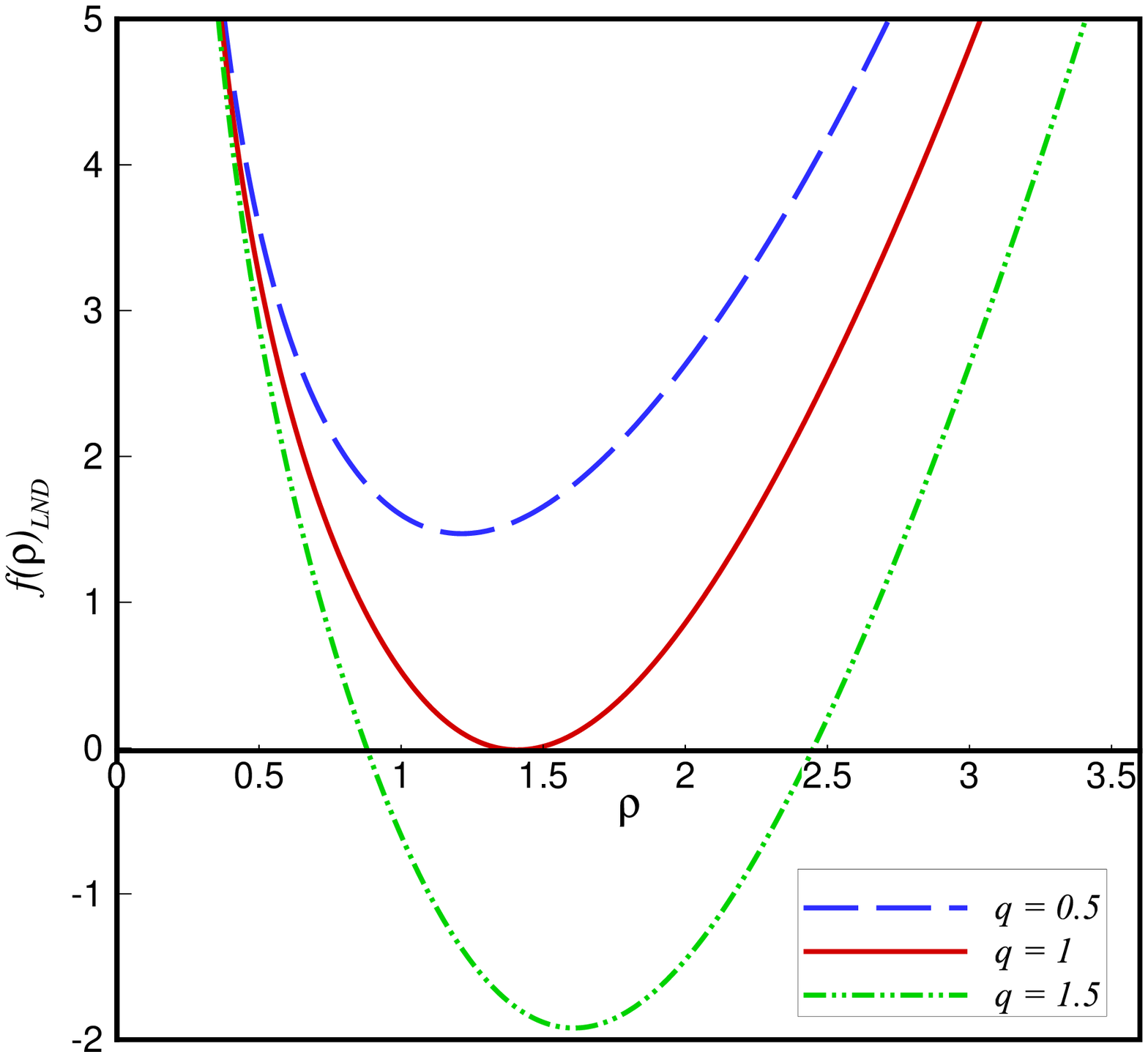}\label{fig3c}}\caption{The behavior
        of $f(\rho)$ versus $\rho$ for $\alpha=2$, $\Lambda=-3$ and $b=1$.}\label{fig3}
\end{figure}
According to these figures, the existence of horizons and their
numbers are crucially depend on the values of the metric
parameters. By suitable choices of the different parameters, one
can see that since $f(\rho)$ is positive near the origin as well as
large distances, the obtained solution may represent black holes
with two horizons, an extreme black hole or naked singularity
depending on the parameters. To be more precise, the effects of
parameters are plotted in different figures. Fig. \ref{fig1}
shows that by fixing other metric parameters, increasing of
$\alpha$  may increase the number of horizons in all presented
theories. From Fig. \ref{fig2}, one may realize that, when other
metric parameters are fixed, a direct relation between the
increase of the nonlinear parameter $\beta$ and the number of
horizon stands for all three metric functions. Also, by fixing
other parameters, the same behavior is seen for electric charge
$q$ in Fig. \ref{fig3}. It is worth noting that the effect of
changes in metric parameters on the number of horizons of
BID-(A)dS, END-(A)dS and LND-(A)dS black holes is quite contrary
to the their counterparts in non-(A)dS spacetime. As mentioned in
\cite{Shey2,ShKa,ShN} for non-(A)dS dilaton black holes with
nonlinear electrodynamics, by fixing other metric parameters,
decreasing of the electric charge parameter $q$, or nonlinear
parameter $\beta$ or dilaton coupling constant $\alpha$ may
increase the number of horizon, but our solutions treat
differently in this regard.
%%%%%%%%%%%%%%%%%%%%%%%%%%%%%%%%%%%%%%%%%%%%%%%%%%%%%%%%%%%%%%%%%%%%%%%%%%%%%%%%%%%%%%%%%%%
\section{Closing remarks}\label{sumsec}
Till now exact analytical solutions of Einstein-dilaton gravity in
the presence of nonlinear electrodynamics and in the background of
(A)dS spaces have not been constructed. In this paper, we
constructed three classes of asymptotically AdS dilaton black
holes solutions in the presence of nonlinear electrodynamics. We
have considered three type of nonlinear electrodynamics, namely
BID, END  and LND field. For this purpose, we utilized a
combination of three Liouville-type dilaton potential. We found
the metric functions, electric field and dilaton field and studied
their behaviour for three class of solutions. We observed that the
nonlinearity of the gauge field do not affect the dilaton field,
and for $r\rightarrow \infty$ ($\rho\rightarrow \infty$), the
dilaton field goes to zero. We proved that our solution has a
singularity at $\rho=0$ ($r=b$), which depending on the value of
the metric function, it can be covered by the one or two horizon.
Also, we realized that the number of horizon increases by
increasing nonlinear parameter $\beta$, electric charge $q$ or
dilaton coupling constant $\alpha$. We studied the behavior of the
electric field of all three nonlinear electrodynamics.
Interestingly, we found out that in the presence of the dilaton
field, the electric field has a strange behaviour. Indeed, it
vanishes at $\rho=0$ and increases until reaches a maximum value
at $\rho=\rho_{0}$ and goes to zero smoothly as
$r\rightarrow\infty$ {($\rho\rightarrow \infty$)}. We observed
that the electric field is zero at singularity and increases
smoothly until reaches a maximum value, then it decreases smoothly
until goes to zero as $r\rightarrow \infty$ {($\rho\rightarrow
\infty$)}. The maximum value of the electric field increases with
increasing the nonlinear parameter $\beta$ or decreasing the
dilaton coupling $\alpha$ and is shifted to the origin in the
absence of either dilaton field ($\alpha=0$) or nonlinear gauge
field ($\beta\rightarrow\infty$). It seems that the coupling of
the nonlinear gauge field with the dilaton field in the background
of (A)dS space leads to such an unfamiliar behavior. However, the
physical reason for this behaviour  and its origin deserves
further investigations and we leave them for the future works. It
would be also interesting to carry out the investigation on the
thermodynamics of the obtained solutions. One may also interested
in generalizing these four dimensional static and spherically
symmetric solutions to the rotating and higher dimensional cases
with other horizon topology. We leave these issues for future
works.
%%%%%%%%%%%%%%%%%%%%%%%%%%%%%%%%%%%%%%%%%%%%%%%%%%%%%%%%%%%%%%%%=%%%%%%%%%%%%%
\acknowledgments{We thank Shiraz University Research Council. This
work has been supported financially by Research Institute for
Astronomy and Astrophysics of Maragha, Iran.}
%%%%%%%%%%%%%%%%%%%%%%%%%%%%%%%%%%%%%%%%%%%%%%%%%%%%%%%%%%%%%%%%%%

\end{document}